\documentclass[%
 reprint,
superscriptaddress,
 amsmath,amssymb,
prb,
]{revtex4-2}

\usepackage{graphicx}
\usepackage{dcolumn}
\usepackage{bm}
\usepackage{appendix}
\usepackage{soul}
\usepackage[utf8]{inputenc}
\usepackage[english]{babel}
\usepackage[T1]{fontenc}
\usepackage[colorlinks,bookmarks=false,citecolor=blue,linkcolor=red,urlcolor=blue]{hyperref}
\usepackage[dvipsnames]{xcolor}
\usepackage{times}
\usepackage{bbm}
\usepackage[normalem]{ulem}
\usepackage{multirow}
\usepackage{booktabs}
\usepackage{amssymb}
\usepackage{amsmath}

\usepackage[final]{changes}

\definechangesauthor[name = JF, color=blue]{JF}
\definechangesauthor[name = GK, color=red]{GK}
\definechangesauthor[name = JP, color=orange]{JP}

\newcommand{\be}{\begin{eqnarray}}
\newcommand{\ee}{\end{eqnarray}}

\begin{document}
\preprint{APS/123-QED}

\title{ Charge-density-wave transitions, phase diagram, soft phonon and possible electronic nematicity:  a thermodynamic investigation of BaNi$_2$(As,P)$_2$}

\author{Christoph~Meingast}
\affiliation{Institute for Quantum Materials and Technologies, Karlsruhe Institute of Technology, 76021 Karlsruhe, Germany}

\author{Anmol~Shukla}
\affiliation{Institute for Quantum Materials and Technologies, Karlsruhe Institute of Technology, 76021 Karlsruhe, Germany}

\author{Liran Wang}
\affiliation{Institute for Quantum Materials and Technologies, Karlsruhe Institute of Technology, 76021 Karlsruhe, Germany}

\author{Rolf~Heid}
\affiliation{Institute for Quantum Materials and Technologies, Karlsruhe Institute of Technology, 76021 Karlsruhe, Germany}

\author{Frédéric~Hardy}
\affiliation{Institute for Quantum Materials and Technologies, Karlsruhe Institute of Technology, 76021 Karlsruhe, Germany}

\author{Mehdi~Frachet}
\affiliation{Institute for Quantum Materials and Technologies, Karlsruhe Institute of Technology, 76021 Karlsruhe, Germany}

\author{Kristin~Willa}
\affiliation{Institute for Quantum Materials and Technologies, Karlsruhe Institute of Technology, 76021 Karlsruhe, Germany}

\author{Tom~Lacmann}
\affiliation{Institute for Quantum Materials and Technologies, Karlsruhe Institute of Technology, 76021 Karlsruhe, Germany}

\author{Matthieu~Le Tacon}
\affiliation{Institute for Quantum Materials and Technologies, Karlsruhe Institute of Technology, 76021 Karlsruhe, Germany}

\author{Michael~Merz}
\affiliation{Institute for Quantum Materials and Technologies, Karlsruhe Institute of Technology, 76021 Karlsruhe, Germany}
\affiliation{Karlsruhe Nano Micro Facility (KNMFi), Karlsruhe Institute of Technology (KIT),
76344 Eggenstein-Leopoldshafen, Germany}

\author{Amir-Abbas~Haghighirad}
\affiliation{Institute for Quantum Materials and Technologies, Karlsruhe Institute of Technology, 76021 Karlsruhe, Germany}

\author{Thomas~Wolf}
\affiliation{Institute for Quantum Materials and Technologies, Karlsruhe Institute of Technology, 76021 Karlsruhe, Germany}

\date{\today}
 
\begin{abstract}
A detailed investigation of BaNi$_2$(As,P)$_2$ single crystals using high-resolution thermal-expansion, heat-capacity, Young’s-modulus and resistivity measurements is presented. The experimental data are complemented by density-functional calculations.   The phase diagram of BaNi$_2$(As,P)$_2$  is shown to be much richer than suggested by the original data of Kudo et al. [Phys. Rev. Lett. 109, 097002 (2012)].  The transition to the commensurate charge-density-wave (C-CDW) is always preceded by a four-fold symmetry-breaking transition associated with the long-range ordering of a strongly fluctuating unidirectional incommensurate charge-density wave (I-CDW). Significant precursors above the I-CDW and C-CDW transitions are seen in the thermal expansion and resistivity and are particularly evident in the temperature dependence of the c/a ratio of the lattice parameters. Heat-capacity measurements of the crystals with a higher P content and a higher critical temperature of 3.2 K uncover a Debye-like behavior of a soft-phonon mode with a very low $\Theta_{\textrm{Debye}}$ of roughly 15 K.  Associated with this soft phonon are unusually large thermal-expansion anomalies, resulting in logarithmically diverging uniaxial phonon Grüneisen parameters. Young’s-modulus data of these higher-T$_c$ crystals exhibit a significant softening in both B$_\textrm{1g}$ and B$_\textrm{2g}$ channels, which is argued to be incompatible with nematic criticality and is rather associated with a broad phase transition to an hitherto unknown structure. Possible origins of the increase in the superconducting critical temperature with P-substitution are discussed.

\end{abstract}

\maketitle
\maketitle


\section{\label{sec:level1}introduction}

BaNi$_2$As$_2$ crystallizes in the same ThCr$_2$Si$_2$ structure as BaFe$_2$As$_2$~\cite{PfistererNagorsen1980}, and, interestingly, undergoes a structural phase transition to a triclininc phase at a similar temperature (135 K) as the stripe spin-density-wave (SDW) transition in BaFe$_2$As$_2$~\cite{Ronning_2008,SefatPRB2009}. However, no signatures of magnetism have been found in either neutron-diffraction ~\cite{Kothapalli} or photoemission ~\cite{zhouPRB2011}  studies. Rather, recent diffraction experiments have shown that the triclinic phase is associated with a commensurate charge-density-wave (C-CDW) order, and that an incommensurate charge-density-wave (I-CDW) precedes the triclinic transition by a few Kelvin ~\cite{EckergPRB2018,LeePRL2019,LeePRL2021,MerzPRB2021}.  Recent high-resolution thermal-expansion data exhibit a clear four-fold symmetry breaking at this I-CDW transition, to an orthorhombic structure with space group $Immm$ ~\cite{MerzPRB2021}. Pure BaNi$_2$As$_2$, in contrast to BaFe$_2$As$_2$, is already superconducting with a very low critical temperature, T$_c$, of 0.7 K ~\cite{Ronning_2008}. T$_c$ can however be increased significantly to roughly 3 K by suppressing the triclinic phase transition through various substitutions on all three atomic positions ~\cite{KudoPRL2012,EckergNP2020,EckergPRB2018}.  In the system BaNi$_2$(As,P)$_2$, Kudo et al. attributed this increase in T$_c$ to an enhanced electron-phonon coupling due to a soft phonon mode, although, surprisingly, no increase in the electronic density of states was found ~\cite{KudoPRL2012}.   In contrast, Eckberg et al. provided evidence for significant B$_\textrm{1g}$ nematic fluctuations via elastoresistive measurements in (Ba,Sr)Ni$_2$As$_2$ and suggested that nematic fluctuations might be the origin of the increased T$_c$ ~\cite{EckergNP2020}.  Theoretically, the phase transitions and superconductivity have been associated with an orbital-selective Peierls instability ~\cite{NodaJPSJ2017}, or orbital fluctuations~\cite{YamakawaJPSJ2013}.  Initial density-functional-theory (DFT) calculations on BaNi$_2$As$_2$ in the tetragonal ThCr$_2$Si$_2$ structure by Subedi and Singh ~\cite{SubediPRB2008} classified this material as a conventional phonon-mediated superconductor due to low-lying optical phonons. Interestingly, the calculated T$_c$ of 4K in this study is close to the highest T$_c$ so far obtained by suppressing the triclinic transition by the various substitutions.  Several questions emerge from these studies. Is T$_c$ raised by soft phonons or by B$_{1g}$ nematic fluctuations, or, alternatively, is there a close connection between these?  Should one even think of an enhanced T$_c$ by these effects, or is it more appropriate to think of a suppressed T$_c$ due to the CDW order and associated distortions in the triclinic phase?  What is the nature of the soft phonon and/or nematic fluctuations in these systems, and how are these influenced by various substitutions?

In this paper we take another look at the BaNi$_2$(As,P)$_2$  system using detailed high-resolution thermal-expansion, heat-capacity, Young’s-modulus and resistivity measurements.  First, the phase diagram of BaNi$_2$(As,P)$_2$ is shown to be much richer than suggested by the original data of Kudo et al.\cite{KudoPRL2012}. In particular, we demonstrate that the triclinic phase is always preceded by a four-fold symmetry-breaking transition, which we attribute to the long-range ordering of the unidirectional I-CDW. Further, crystals with the higher-T$_c$ phase, i.e. in which the triclinic phase is suppressed with P substitution, surprisingly also exhibit a clear phase transition to an unknown low-temperature structure. Significant precursors above the I-CDW and C-CDW transitions are resolved in thermal-expansion and resistivity data and are particularly evident in the temperature dependence of the c/a ratio of the lattice parameters.  The thermal expansion data allow us to study the temperature evolution of the lattice parameters in the tetragonal ThCr$_2$Si$_2$ structural setting with P-substitution. We find that the largest effect occurs perpendicular to the Ni planes, where the lattice shrinks by roughly 1.5$\%$ upon cooling in BaNi$_2$As$_2$.  This c-axis shrinking decreases strongly with increasing P-content and, surprisingly, changes sign to a small expansion for the higher-T$_c$ material. Our detailed heat-capacity measurements uncover a Debye-like behavior of the soft-phonon mode with a very low Debye temperature, T$_{\textrm{Debye}}$,
of roughly 15 K.  Associated with this soft phonon are large thermal-expansion anomalies, resulting in unprecedentedly diverging phonon Grüneisen parameters.  To check for the expected softening of the shear modulus at a nematic phase transition, Young’s-modulus measurements ~\cite{BohmerPRL2015} were performed on the higher-T$_{\textrm{c}}$ phase. A significant softening in both B$_\textrm{1g}$ and B$_\textrm{2g}$ channels is found, and this is argued not to be compatible with nematic criticality. Finally, detailed resistivity data provide evidence for a dominant electron-electron scattering only in the higher-T$_c$ phase.

\section{\label{sec:level1}Methods}

Single crystals of BaNi$_{\textrm{2}}$(As$_\textrm{1-x}$P$_\textrm{x}$)$_{\textrm{2}}$ were grown using a self-flux method, as described in more detail in References~\cite{MerzPRB2021,Yao2022,Fracchet2022}. Electron micro probe analysis of the crystals was performed using a benchtop scanning electron microscope (SEM) – energy dispersive x-ray spectroscopy (EDS) device COXEM EM-30plus equipped with an Oxford detector (operating with AztecLive software). The EDS analyses on the BaNi$_{\textrm{2}}$(As$_\textrm{1-x}$P$_\textrm{x}$)$_{\textrm{2}}$ crystals of the present study revealed phosphorus contents x = 0.035, 0.048, 0.075 and 0.10. Samples from the same batches were also characterized with single crystal x-ray diffraction, confirming the above compositions within the errors of both techniques. Thermal expansion and Young's modulus were measured using a home-built high-resolution capacitance dilatometer~\cite{ChristophPRB1990,BohmerPRL2014}.  A three-point-bending technique ~\cite{BohmerPRL2014} was employed for the Young's modulus determination.  Heat-capacity and resistivity measurements were made in a Physical Property Measurement System (PPMS) from Quantum Design. Details of density-functional calculations (DFT) are described Ref.~\cite{Pokharel2022}.

\section{\label{sec:level1}Results}

In this section we concentrate on making a detailed comparison between crystals with widely different  P contents  - i.e. one with a clear triclinic phase transition and a low T$_\textrm{c} = 0.7 K $ (x = 0.048)  and one without the triclinic transition and a higher T$_\textrm{c}$ = 3.2 K (x = 0.10) - using thermal-expansion, heat-capacity, Young's-modulus and resistivity measurements.  With this comparison we hope to shed new light on the CDW transitions, the soft phonon effects, as well as possible electronic nematicity.  Additionally, the complete phase diagram will be mapped out using thermal expansion measurements on crystals with additional P-contents.

\subsection{Thermal expansion: x = 0.048}

\begin{figure}[!t]
\centering
\begin{minipage}{1\linewidth}
\centering
\includegraphics[width=1.5\linewidth]{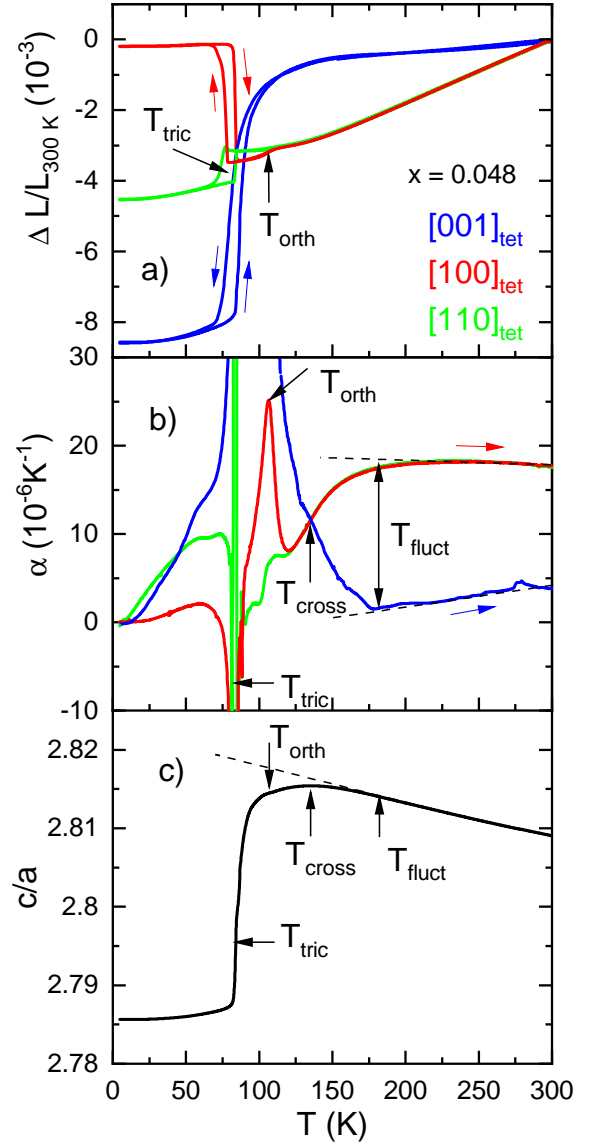}
\end{minipage}


\caption{a) Relative thermal expansion,
$\Delta$ $L/L_\textrm{300K}$, b) thermal- expansion coefficient, $\alpha$ = $1/L\cdot dL/dT$, and c) c/a ratio of the lattice parameters for x = 0.048.  Clearly visible are the four-fold symmetry breaking transition at T$_{\textrm{orth}}$ and the triclinic transition at T$_\textrm{tric}$.  Significant precursors to these transitions are seen in the c/a ratio and are marked by their onset, T$_\textrm{fluct}$, and T$_\textrm{cross}$, where $\alpha_a$ and $\alpha_c$ cross leading to the maximum in c/a. The colored arrows indicate cooling and heating curves. (see text for details)}      
\label{Fig1-Thermalexpansion}
\end{figure}

In Figure~\ref{Fig1-Thermalexpansion} the relative length changes, $\Delta$ $L/L_\textrm{300K}$, along the tetragonal $\left[100 \right]_{\textrm{tet}}$, $\left[ 110 \right]_{\textrm{tet}}$ and $\left[ 001 \right]_{\textrm{tet}}$ directions upon cooling and warming (Fig.~\ref{Fig1-Thermalexpansion}a) are plotted for a crystal with x = 0.048, together with the corresponding thermal expansion coefficients, $\alpha$ = $1/L\cdot dL/dT$, upon warming (Fig.~\ref{Fig1-Thermalexpansion}b).  Clearly observed are both the transitions (upon cooling and heating) to the orthorhombic phase discussed in ~\cite{MerzPRB2021} at T$_{\textrm{orth}}$ = 106 K and to the triclinic phase at T$_{\textrm{tric}}$ = 80 K ~\cite{SefatPRB2009, EckergPRB2018,EckergNP2020,LeePRL2019,LeePRL2021,MerzPRB2021}.  The discontinuous change in lengths and the large thermal hysteresis at T$_{\textrm{tric}}$ are consistent with the strong first-order character of the triclinic transition. Hereafer, We will continue to use the tetragonal notation for the triclinic phase, since it is impossible to reorient the crystal along the triclinic axes in the dilatometer. Thus, although the thermal expansion below T$_{\textrm{tric}}$ does not correspond to the triclinic crystallographic directions, this, as will be shown, provides very useful information about the general structural evolution with temperature for different P-substitutions.  Moreover, the triclinic distortion is quite small with respect to the tetragonal room-temperature structure ~\cite{SefatPRB2009,KudoPRL2012,EckergPRB2018,EckergNP2020,LeePRL2019,LeePRL2021,MerzPRB2021}. The largest thermal-expansion effect occurs along the $\left[ 001 \right]_{\textrm{tet}}$ direction, i.e. perpendicular to the NiAs planes, where the lattice shrinks by about 0.85$\%$ upon cooling. This shrinking is not restricted to the triclinic phase transition itself, but already starts significantly above T$_{\textrm{tric}}$, as clearly demonstrated by the expansivity in Fig.~\ref{Fig1-Thermalexpansion}b.  Here we can identify two related temperatures, 1) T$_{\textrm{cross}}$ where the in-plane and out-of-plane thermal-expansion coefficients cross and 2) T$_{\textrm{fluct}}$ which marks the onset of the transition, i.e. where the expansivities start to deviate from a more regular temperature dependence.  This behavior is mirrored nicely in the $c/a$ ratio shown in Figure~\ref{Fig1-Thermalexpansion}c, calculated using room-temperature values of the lattice parameters from Refs.  \cite{KudoPRL2012,MerzPRB2021}. Since the expansivity at high temperatures along $\left[ 001 \right]_{\textrm{tet}}$ is much smaller than along $\left[ 100 \right]_{\textrm{tet}}$ , the $c$/$a$ ratio increases with decreasing temperatures until reaching a maximum at T$_{\textrm{cross}}$ and then decreases sharply at the triclinic transition.  The temperatures T$_{\textrm{fluct}}$, T$_{\textrm{cross}}$, T$_{\textrm{orth}}$ and T$_{\textrm{tric}}$ will later be used for mapping out the phase diagram.  It is also worth noting that the expansivity at 300 K is already anomalous, since $\alpha_{\textrm{$\left[ 001 \right]$}}$ is much smaller than $\alpha_{\textrm{$\left[ 100 \right]$}}$ (see Figure~\ref{Fig1-Thermalexpansion}b).  This is opposite of what is found for BaFe$_2$As$_2$ ~\cite{Meingast2012PRL} and is an indication for additional physics besides the normal anharmonic phonon-induced thermal expansion operating already in the high-temperature tetragonal phase. 

\begin{figure}[!t]
\centering
\begin{minipage}{1.0\linewidth}
\centering
\includegraphics[width=1.0\linewidth]{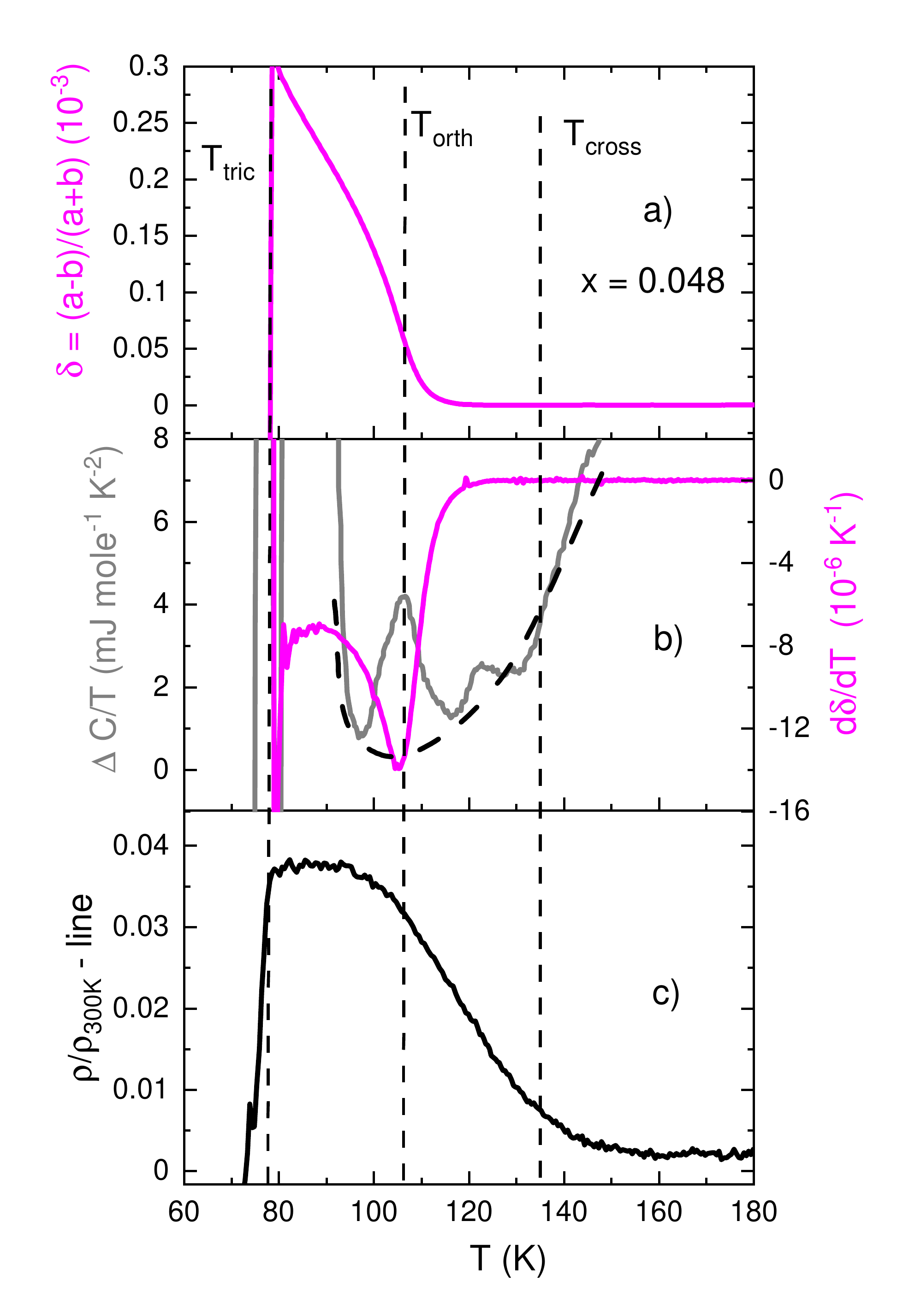}
\end{minipage}
\caption{Temperature dependence of a) orthorhombic distortion, $\delta$, b) heat capacity and d$\delta$/dT and c) in-plane resistivity, $\Delta \rho$/$\rho _\textrm{300K}$ around T$_\textrm{orth}$ for x = 0.048. Whereas clear anomalies are observed in both $\Delta$C$_p$ and d$\delta$/dT, no clear anomaly is seen in the resistivity. Linear backgrounds have been subtracted from both heat-capacity and resistivity data.  (see text for details)}
\label{Figure2}
\end{figure}

\subsection{Four-fold symmetry-breaking transition at T$_{\textrm{orth}}$: x = 0.048}

As demonstrated previously ~\cite{Wang2019JPSJ,BohmerNC2015,LiranPRB2016,MerzPRB2021}, an orthorhombic distortion can be detected in a capacitance dilatometer by measuring the expansion along two different in-plane directions due to the detwinning effect of the small uniaxial pressure applied by the device.  In Fig~\ref{Fig1-Thermalexpansion}a, a clear splitting of the thermal expansion along $\left[ 100 \right]_{\textrm{tet}}$ and $\left[ 110 \right]_{\textrm{tet}}$ directions is observed below T$_{\textrm{orth}}$, marking the onset of in-plane symmetry breaking.  Here the ‘detwinning’ effect occurs for pressure applied along the $\left[ 100 \right]_{\textrm{tet}}$ direction, suggesting a transition to an orthorhombic structure with space group $Immm$ ~\cite{MerzPRB2021}.  Together the $\left[ 100 \right]_{\textrm{tet}}$ and $\left[ 110 \right]_{\textrm{tet}}$ data sets allow (under the assumption that the uniaxial pressure applied by the dilatometer is large enough to fully detwin the crystal) to calculate the orthorhombic distortion  $\delta$=(a-b)/(a+b) - see Figure~\ref{Figure2}a.  Note that the orthorhombic distortion in BaFe$_{2}$As$_{2}$ is significantly larger and occurs along the $\left[ 110 \right]_{\textrm{tet}}$ instead of the $\left[ 100 \right]_{\textrm{tet}}$ direction in the Ni compound ~\cite{Rotter2008PRB}.  In Fig.~\ref{Figure2}b the derivative of $\delta$ is plotted together with heat-capacity data, for which a straight line has been subtracted in order to make the tiny anomaly, which is only about $\Delta$C$_p/T$ = 4 mJ/(mole K$^2$), visible.  Both quantities exhibit a sharp peak at T$_{\textrm{orth}}$ = 106 K.  This behavior is reminiscent of a strongly fluctuating low-dimensional transition, in contrast to the very mean-field-like behavior in Fe-based nematic transitions ~\cite{BohmerNC2015,BohmerPRL2015,ChuPRB2009,DeJonghAdvainPhyP2001}.  Surprisingly, no clear feature is seen in the in-plane resistivity at T$_{\textrm{orth}}$ (note - here a straight line has been subtracted to highlight the anomaly) - rather a very broad increase in the resistivity starting at roughly 150 K is observed, which suggests a correlation of the resistivity anomaly with T$_{\textrm{fluct}}$ and T$_{\textrm{cross}}$ instead of T$_{\textrm{orth}}$. This interpretation is also supported by the resistivity data for x = 0.1 (see Fig.~\ref{Fig9} and other detailed resistivity measurements in Ref.~\cite{Fracchet2022}).

\subsection{Thermal expansion x = 0.10: evidence for an additional phase transition}

\begin{figure}[!t]
\centering
\begin{minipage}{1.0\linewidth}
\centering
\includegraphics[width=1.5\linewidth]{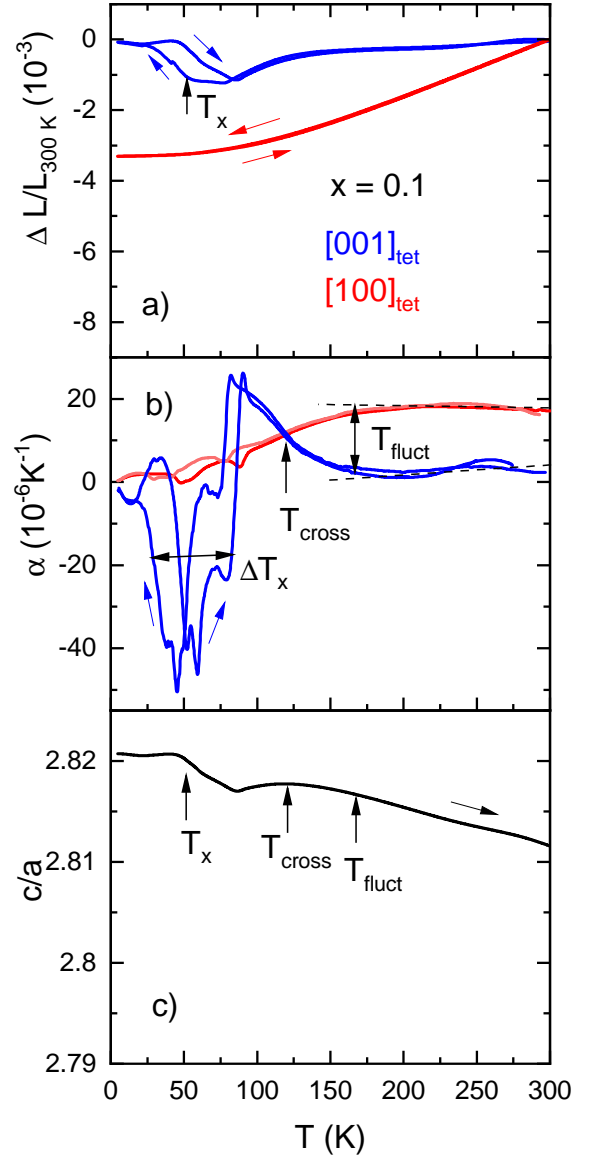}
\end{minipage}
\caption{ a) Relative uniaxial thermal expansion, $\Delta  L/L_\textrm{300K}$, b) thermal-expansion coefficients, $\alpha_{\textrm{100}}$ and $\alpha_{\textrm{001}}$, and c) c/a ratio of the lattice parameters for x = 0.1.  Clearly visible along the $\left[ 001 \right]_{\textrm{tet}}$ direction is a  first-order phase transition at T$_x$ to an hitherto unknown structure. The colored arrows indicate cooling and heating curves. In contrast to the c/a ratio for x = 0.048, c/a does not exhibit a large reduction upon cooling. (see text for details)}
\label{Figure3}
\end{figure}

Next we turn to the thermal expansion of the crystal with x = 0.1, which is not expected to exhibit a triclinic transition ~\cite{KudoPRL2012}.  Surprisingly, clear signs of a phase transition of unknown origin for this composition are observed in the thermal expansion. In particular, the thermal expansion along the $\left[ 001 \right]_{\textrm{tet}}$ direction exhibits an expansion upon cooling at a strongly hysteretic transition at T$_x$ = 50 K (see Fig.~\ref{Figure3}a).  This expansion counteracts the contraction due to high-temperature fluctuations, such that the total relative length change along the c-axis from 300 K to 5 K is nearly zero.  In contrast, the transition at T$_x$  has only a very small effect upon the in-plane lengths, and is made only visible  in the $\left[ 100 \right]_{\textrm{tet}}$ expansion coefficients (see Fig.~\ref{Figure3}b). We see no sign of the orthorhombic transition, at which one would expect a positive signal in $\alpha_{\textrm{100}}$ (see Fig. ~\ref{Fig1-Thermalexpansion}b) due to the detwinning effect of the dilatometer, which always measures the shorter of the two orthorhombic axes. Instead, $\alpha_{\textrm{100}}$ drops slightly at the onset at T$_x$ (see red lines in Fig. ~\ref{Figure3}b) and also exhibits a significant hysteresis.  Both of these findings are inconsistent with the second-order orthorhombic transition seen in all P concentrations ~\cite{MerzPRB2021}. Possible origins of this transition will be discussed further on. At higher temperatures we can again identify both T$_{\textrm{cross}}$ and T$_{\textrm{fluct}}$ as in the crystals with lower P content.  Both of these temperatures are slightly reduced in comparison to those found for x = 0.048.

 \subsection{c-axis thermal expansion and structural evolution with P-substitution}

\begin{figure}[!t]
\centering
\begin{minipage}{1.0\linewidth}
\centering
\includegraphics[width=1.2
\linewidth]{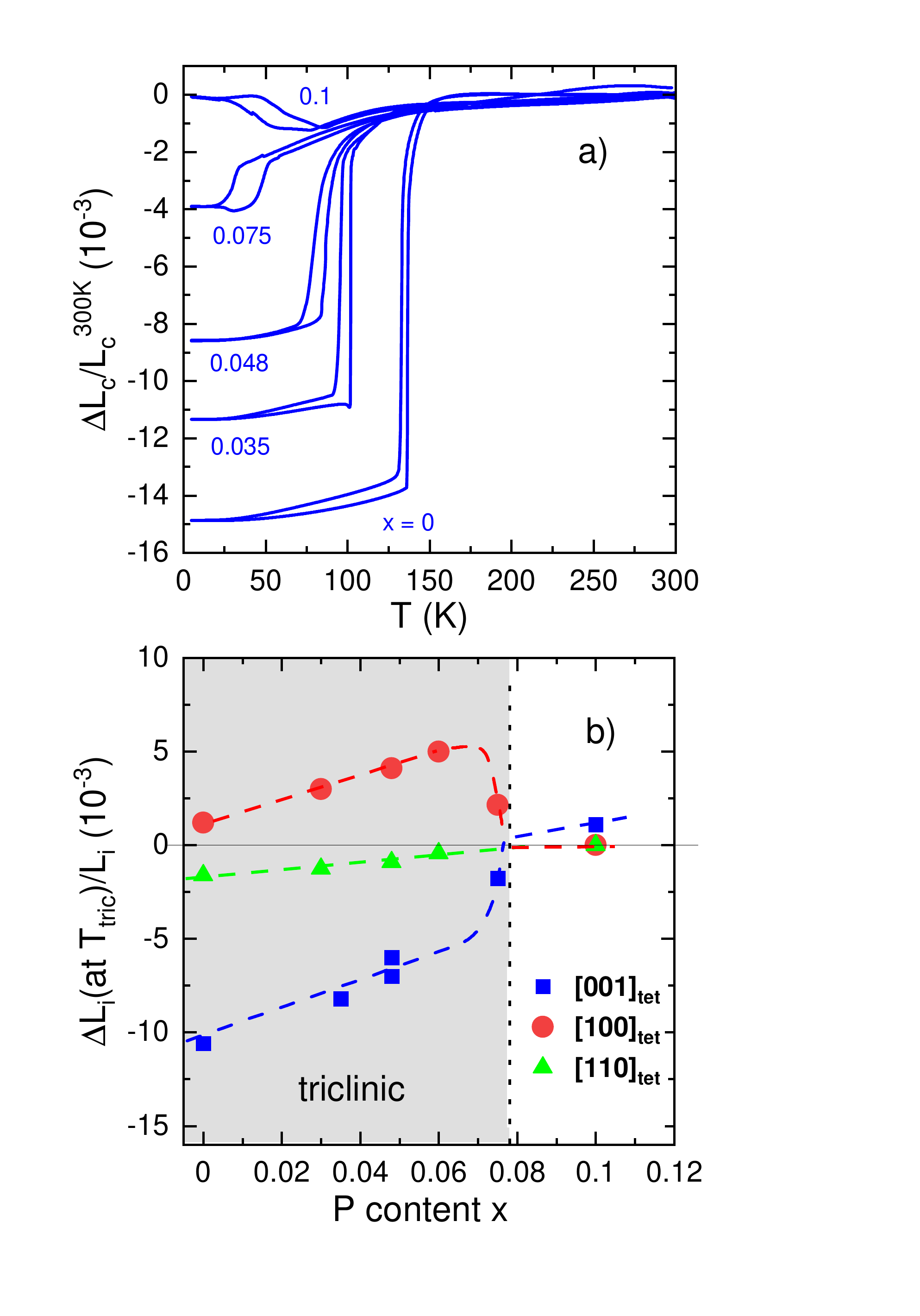}
\end{minipage}
\caption{ a) Temperature dependence of the relative thermal expansion along $\left[ 001 \right]_{\textrm{tet}}$, demonstrating the suppression of the c-axis contraction and the sign change upon P - substitution. b) P-content dependence of the relative length changes at the transitions T$_{\textrm{tric}}$ and T$_x$ for all three tetragonal directions. (see text for details)  }
\label{Fig4}
\end{figure}

The evolution of the c-axis thermal expansion for several additional P-substitutions is shown in Fig.~\ref{Fig4}.  The magnitude of the c-axis contraction upon cooling is drastically reduced with increasing P content and changes sign to a small expansion for x = 0.1. The opposite sign of the transition for x = 0.1, as well as the fact that it occurs at a higher temperature than the triclinic transition for x = 0.075, strongly imply that this is not due to an inhomogeneous crystal with a remaining fraction of the triclinic phase, but rather to a new phase. The P-dependence of the length changes along all three measured directions at T$_{\textrm{tric}}$ and T$_x$ for x = 0.10 are presented in Fig.~\ref{Fig4}b for all investigated compositions.  Whereas the magnitude of the shrinking along $\left[ 001 \right]_{\textrm{tet}}$ decreases with increasing x, the expansion along $\left[ 100 \right]_{\textrm{tet}}$ increases with x until about x = 0.075, beyond which all effects are abruptly reduced in magnitude. This highlights the first-order nature of the vanishing triclinic phase with P-substitution.

\subsection{Phase diagram}

\begin{figure}[!t]
\centering
\begin{minipage}{1.0\linewidth}
\centering
\includegraphics[width=1.2\linewidth]{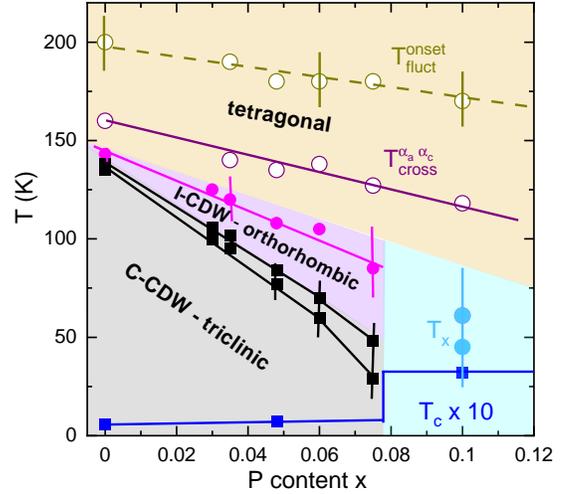}
\end{minipage}
\caption{Phase diagram of BaNi$_2$(As,P)$_2$ determined from thermal-expansion and heat-capacity data. Closed symbols represent real thermodynamic phase transitions, and open symbols represent crossover temperatures. T$_{\textrm{fluc}}$ and T$_{\textrm{cross}}$ decrease monotonically across the whole doping range. In contrast, T$_{\textrm{orth}}$ and T$_{\textrm{tric}}$ disappear near x = 0.08 and are replaced by the transition at T$_x$, leading to the higher superconducting T$_c$.  The thermal hysteresis for cooling and heating at T$_{\textrm{tric}}$ and T$_x$ is indicated by double symbols. Closed symbols represent real phase transitions, and open symbols refer to cross-over temperatures. (see text for details)}
\label{Fig5}
\end{figure}

The thermal-expansion data allows us to construct a detailed phase diagram showing the evolution of  T$_{\textrm{fluct}}$, T$_{\textrm{cross}}$, T$_{\textrm{orth}}$, T$_{\textrm{tric}}$ with P substitution, as well as the newly discovered transition at T$_x$ (see Fig.~\ref{Fig5}. T$_{\textrm{fluc}}$ and T$_{\textrm{cross}}$ decrease slightly with x and are visible over the entire range of x studied, suggesting that the fluctuations at high temperature are present over this whole substitution region.  On the other hand, T$_{orth}$ and T$_{\textrm{tric}}$ decrease significantly faster and are only observed up to x = 0.075. The higher T$_{c}$ occurs only in the higher substitution region (i.e. x > 0.075), as detailed by the original work of Kudo et al. \cite{KudoPRL2012}.  The x-dependence of T$_{\textrm{orth}}$, the high-temperature fluctuations characterized by T$_{\textrm{fluct}}$ and T$_{\textrm{cross}}$, as well as the transition at T$_x$ are new findings, which were previously ~\cite{KudoPRL2012} not recognized in the phase diagram of BaNi$_2$(As,P)$_2$.

\subsection{Heat capacity for x = 0.048 and x = 0.1: evolution of soft phonon}

\begin{figure}[!t]
\centering
\begin{minipage}{1.0\linewidth}
\centering
\includegraphics[width=1.1\linewidth]{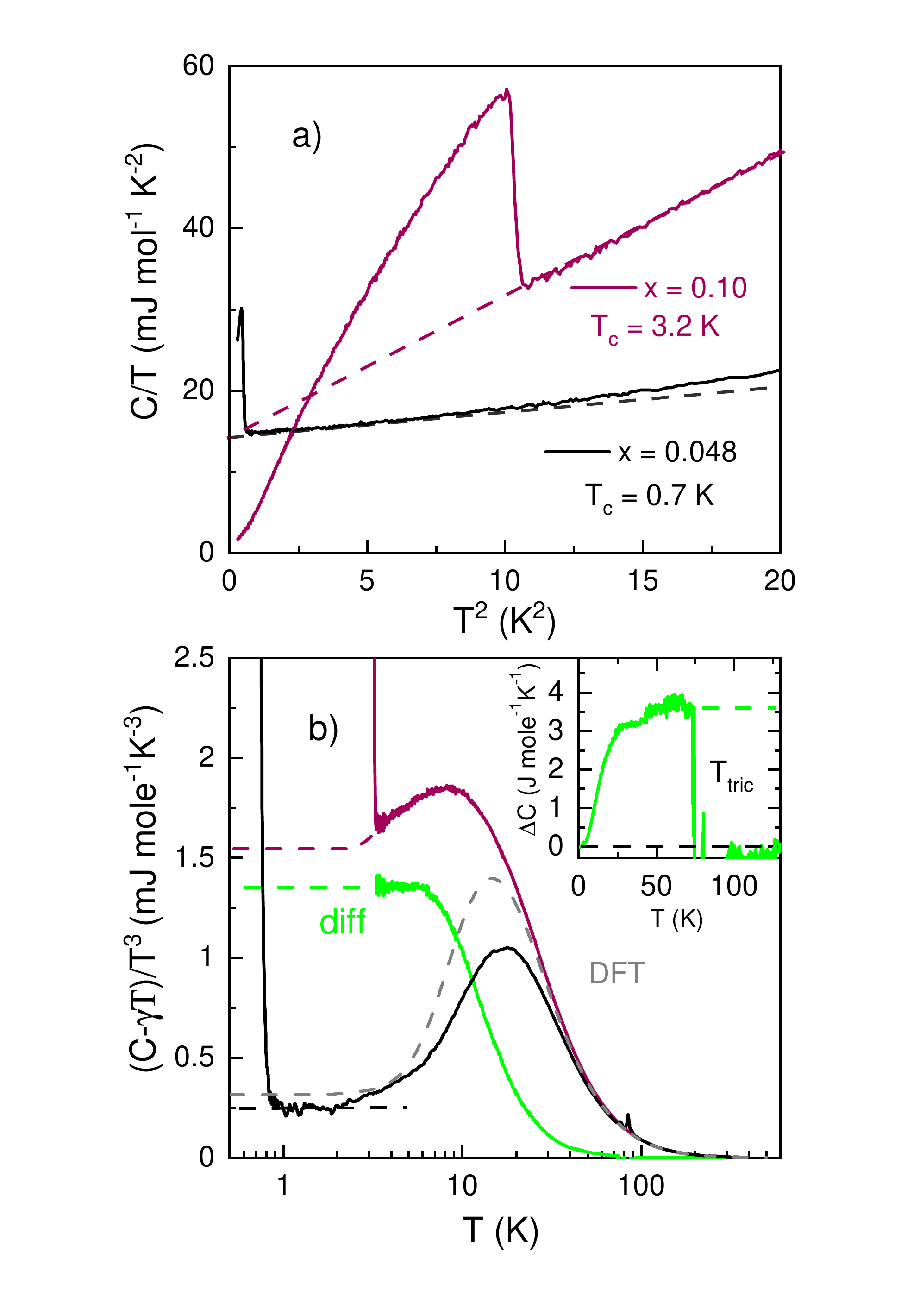}
\end{minipage}
\caption{ a) Heat capacity C/T versus T$^{2}$ for x = 0.048 and 0.1. Dashed lines are linear fits demonstrating the increase in the phonon Debye term with increasing P content, as well as the  P-content independent electronic $\gamma$. b) Phonon heat capacity divided by T$^{3}$.  Here, the low-T constant value is indicative of Debye behavior due to acoustic phonons, and the peaks near 10 - 20 K result from optical phonons.  The difference between x = 0.1 and x = 0.048 (green curve) is reminiscent of a Debye-like heat capacity with a very low $\theta_\textrm{Debye}$ = 10 - 15 K and disappears above the triclinic transition (see inset). Heat capacity from DFT calculations in the stable tetragonal state for x = 0 (dashed gray curve) is very similar to the experimental data for x = 0.048. }
\label{Fig6}
\end{figure}

In the following we further analyze our data in order to gain a better understanding of the large increase of T$_c$ from pure BaNi$_2$As$_2$ to crystals with x = 0.1.  A comparison of the heat capacity of crystals with x = 0.048 and x = 0.1 are shown in Fig.~\ref{Fig6} in a C$_{p}$/T versus T$^{2}$ representation.  As observed previously~\cite{KudoPRL2012}, the crystal with the higher T$_c$ exhibits a much larger slope, i.e. Debye contribution, suggesting a significant lattice softening of the acoustic modes.  On the other hand, the Sommerfeld coefficient, $\gamma$, is surprisingly independent of the P-content, in agreement with Ref. ~\cite{KudoPRL2012}.  By subtracting the electronic term $\gamma$T from C$_{p}$ and plotting the resulting phonon heat capacity divided by T$^3$ vs logT, we can obtain more detailed information about the evolution of the anomalous Debye (T$^{3}$) term (see Fig.~\ref{Fig6}b).  In this representation, a Debye heat-capacity gives a constant C$_{\textrm{phonon}}$/T$^{3}$ term well below the Debye temperature, T$_{\textrm{Debye}}$, and then smoothly approaches zero at higher temperatures.  A peak, on the other hand, is indicative of optical phonons, which can be represented by Einstein heat capacities.  The expected Debye behavior for the x = 0.048 crystal is clearly seen at very low T  (between 0.7 and 2.5 K), and low-lying optical phonons give rise to the peak centered at roughly 20 K.  For the x = 0.1 crystal, C$_{\textrm{phonon}}$/T$^{3}$ is significantly larger at low T, but never really reaches a constant value before entering the superconducting state at 3.2 K.  The difference between these curves (green curve) suggests that the large phonon softening observed in the x = 0.1 composition arises from a Debye-like term with an extremely low T$_{\textrm{Debye}}$ of approximately 10-20 K. Further, since this difference vanishes above T$_{\textrm{tric}}$ = 80 K (see inset in Fig.~\ref{Fig6}b), this clearly demonstrates that this phonon softening is already present also in the x = 0.048 crystal above T$_{\textrm{tric}}$ and that it vanishes abruptly at T$_{\textrm{tric}}$. Interestingly, the phonon heat capacity from our DFT calculations in the tetragonal structure (gray dashed curve in Fig.~\ref{Fig6}b) roughly match data for the triclinic phase (x = 0.048) both in the location of the optical phonon peak and the value of the Debye term.  In these calculations, the E$_{g,1}$ phonon instability  was intentionally suppressed by using a coarse energy sampling ~\cite{Pokharel2022}, and this comparison thus suggests that the soft phonon mode is most likely directly related to this instability and is not an intrinsic feature of a stable tetragonal phase.

\subsection{Diverging uniaxial phonon Grüneisen parameters}

\begin{figure}[!t]
\centering
\begin{minipage}{1.0\linewidth}
\centering
\includegraphics[width=1.0\linewidth]{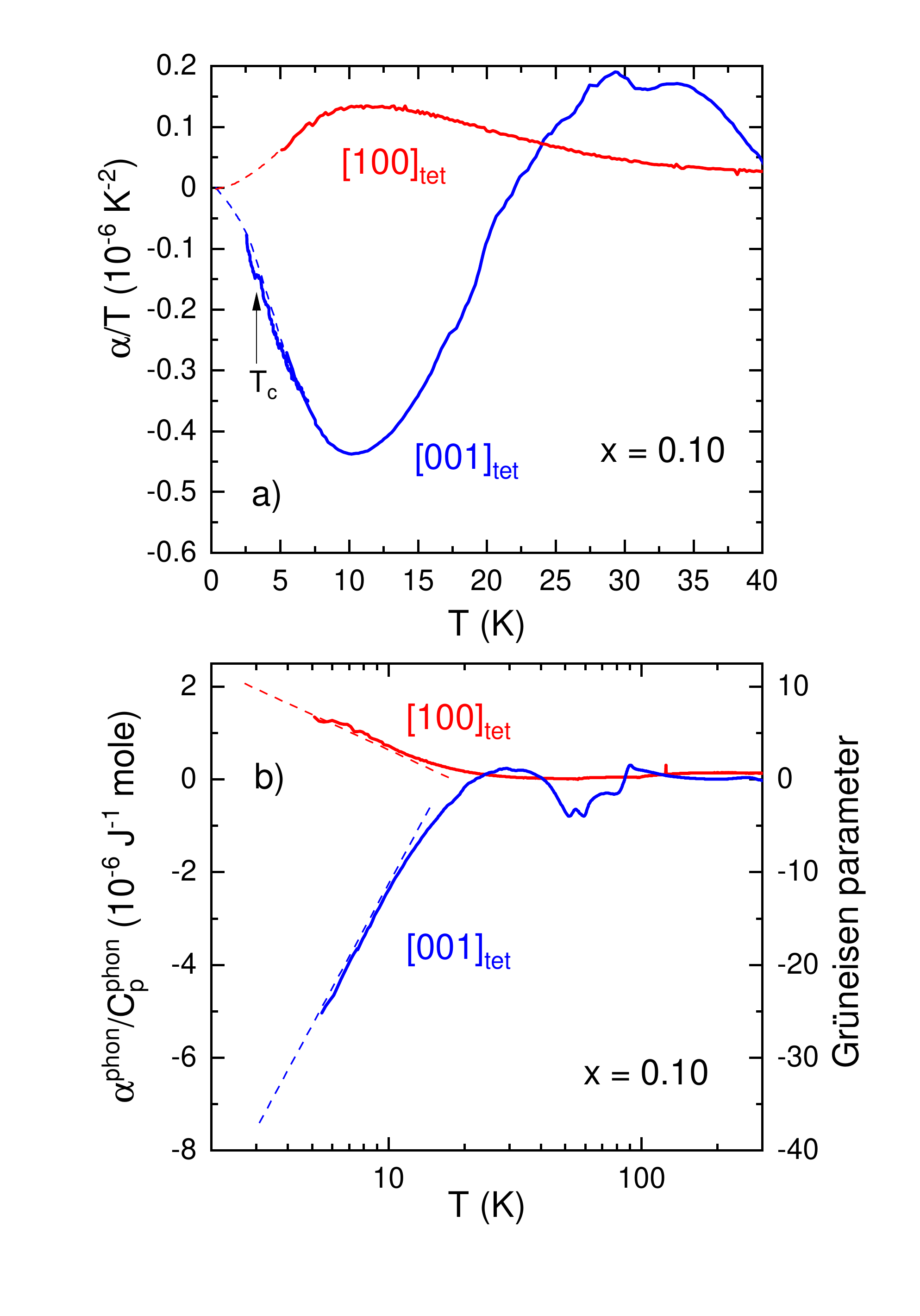}
\end{minipage}
\caption{ a) Maxima and minima in  $\alpha_{\left[ 100 \right]}$/T and  $\alpha_{\left[ 001 \right]}$/T near 10 K for x = 0.1. The dashed lines represent extrapolations to T = 0 K, suggesting that electronic effects, due to the pressure dependence of the Sommerfeld coefficient are small. The superconductivity induced anomaly at T$_c$ is barely visible along $\left[ 001 \right]_{\textrm{tet}}$. b) Phonon uniaxial Grüneisen parameters,  $(\alpha / C_p)^{phon}$, highlighting a logarithmic increase down to low T. The values of $\Gamma^{\textrm{Grüneisen}}$ on the right vertical scale were estimated assuming a 'normal' value of 1.5 at room temperature.  (see text for details)   }
\label{Fig7}
\end{figure}

Additional information about this soft-phonon mode can be obtained by examining the thermal-expansion coefficients and their associated uniaxial Grüneisen-parameters. As shown in Fig.~\ref{Fig7}, the low-temperature thermal expansion for x = 0.1 is quite anomalous.  Here, in a plot of $\alpha$/T vs T a clear maximum and a minimum at around 10 K are observed along $\left[ 100 \right]_{\textrm{tet}}$ and $\left[ 001 \right]_{\textrm{tet}}$ directions, respectively.  An extrapolation of these curves to T = 0 K yield values of $\alpha$/T near zero (dashed lines in Fig.~\ref{Fig7}), which excludes an electronic origin related to the pressure dependence of the Sommerfeld coefficient. This is also consistent with the extremely small anomaly at T$_c$ in $\left[ 001 \right]_{\textrm{tet}}$, since this is a purely electronic effect. This maximum (minimum) is clearly closely related to the larger T$^{3}$ term of the heat capacity, since no such anomalies are observed in the triclinic phase (see Fig.~\ref{Fig1-Thermalexpansion}b).   Now assuming that we have both the phonon thermal expansion and heat capacity, we can calculate the uniaxial phonon Grüneisen parameters.  Without detailed knowledge of the elastic constants, we simply plot the ratio of $\alpha_i$/C$_{\textrm{phonon}}$ in Fig.~\ref{Fig7}b, which should give a good approximation of the temperature dependence of the lattice Grüneisen parameters.  The behavior of $\Gamma^{\textrm{Grüneisen}}$ is quite novel. At low temperature $\Gamma^{\textrm{Grüneisen}}$ along both $\left[ 100 \right]_{\textrm{tet}}$ and $\left[ 001 \right]_{\textrm{tet}}$ directions exhibits a logarithmic divergence down to the lowest temperature measured, as indicated by the dashed lines.  The opposite sign of this divergence along a- and c-axes indicates a close connection with the c/a ratio.  In order to obtain a feeling for the magnitude of these Grüneisen parameters, we assign a typical $\Gamma^{\textrm{Grüneisen}}$ value of 1.5 to the volume Grüneisen at 300 K and then obtain the estimated values shown on the right scale of Fig.~\ref{Fig7}b.  The value of 25 at 5 K along the  $\left[ 001 \right]$ direction is anomalously enhanced for typical phonon anharmonicity. The logarithmic divergence is suggestive of the proximity to a novel phonon quantum critical behavior.  Quantum critical behavior is often associated with a low-temperature accumulation of entropy, for which a sign change of the thermal expansion under an appropriate tuning parameter is a tell-tale sign ~\cite{GarstPRB2005}. In our data we see sign changes of the thermal expansion coefficients both with temperature at T$_{\textrm{cross}}$ (see Fig.~\ref{Fig1-Thermalexpansion}b and Fig.~\ref{Figure3}b) and with doping (see Fig.~\ref{Fig4}a). We are not aware of similar phonon effects in other systems, since quantum criticality is usually associated with electronic degrees of freedom ~\cite{SachdevPhyToday2011,GarstPRB2005}.

\subsection{Young's modulus: nematicity?}

\begin{figure}[!t]
\centering
\begin{minipage}{1.0\linewidth}
\centering
\includegraphics[width=1.2\linewidth]{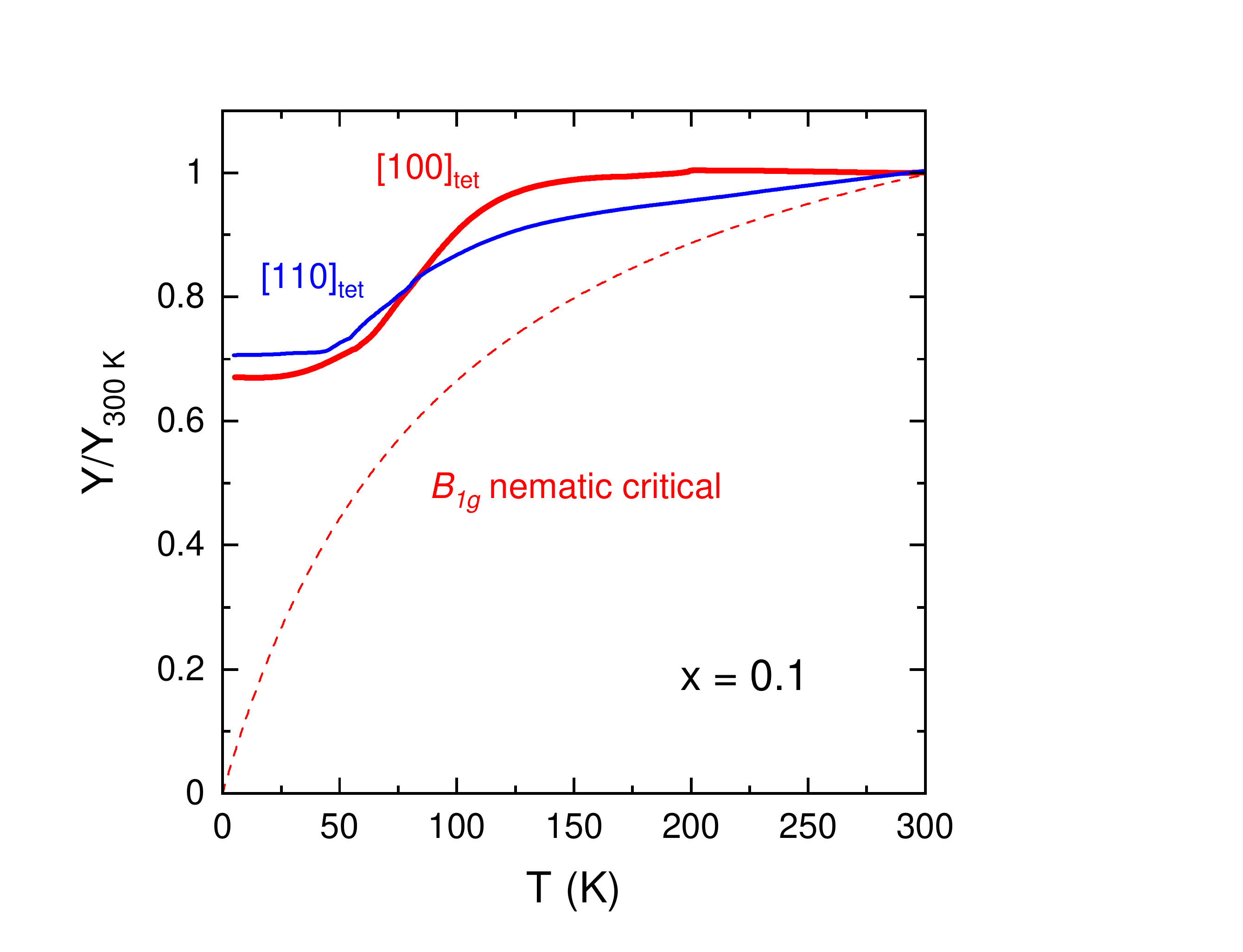}
\end{minipage}
\caption{Normalized Young's modulus, $Y/Y_\textrm{300K}$, for a crystal with x = 0.1 measured using a three-point-bending technique in a capacitance dilatometer. $Y/Y_\textrm{300K}$ softens significantly between 50 K and 100 K along both B$_{1g}$ (red) and B$_{2g}$ (blue) directions, but the data are inconsistent with a hypothetical B$_{1g}$ nematic criticality (red dashed line) in which Y vanishes at T = 0 K.  (see text for details)  }
\label{Fig8}
\end{figure}

In order to search for nematic signatures in the higher-T$_c$ crystal with x = 0.1, Young’s modulus, $Y$, measurements were made using a three-point-bending technique in our capacitance dilatometer ~\cite{BohmerPRL2014,LiranPRB2018}.  We note that $Y$ is, to a good approximation, proportional to the shear modulus near a real nematic transition ~\cite{BohmerPRL2014,LiranPRB2018}.  The temperature dependence of the normalized $Y$, $Y$/$Y_\textrm{300K}$, are plotted in Fig. \ref{Fig8} along both  $\left[ 100 \right]_{\textrm{tet}}$ and  $\left[110 \right]_{\textrm{tet}}$ directions.  Although a significant softening is observed between 50 K and 150 K in both sets of data, the data are not compatible with nematic criticality.  First, the softening is roughly of equal magnitude along both directions.  Second, $Y_\textrm{}$/$Y_\textrm{300K}$ is nearly constant below about 50 K.  The red dashed curve represents a hypothetical nematic critical curve along $\left[100 \right]_{\textrm{tet}}$, i.e. in the B$_\textrm{1g}$ channel.  The constant values of $Y$/$Y_\textrm{300K}$ are reminiscent of the behavior of the Fe-based materials below the magnetic/nematic transitions ~\cite{BohmerPRL2014,BohmerPRL2015}. The temperature range of the present softening corresponds roughly to the anomalous thermal expansion between T$_x$ and T$_\textrm{fluct}$ and not to the temperature of the soft phonon (or anomalous Grüneisen parameters), which happen at much lower temperatures. We note that the lack of a large critical nematic response in $Y$/$Y_\textrm{300K}$ is consistent with elastoresistivity ~\cite{Fracchet2022} and ARPES data under uniaxial stress ~\cite{Guo2022}.

\subsection{Resistivity scaling for x = 0.048 and x = 0.10}

\begin{figure}[!t]
\centering
\begin{minipage}{1.0\linewidth}
\centering
\includegraphics[width=1.2\linewidth]{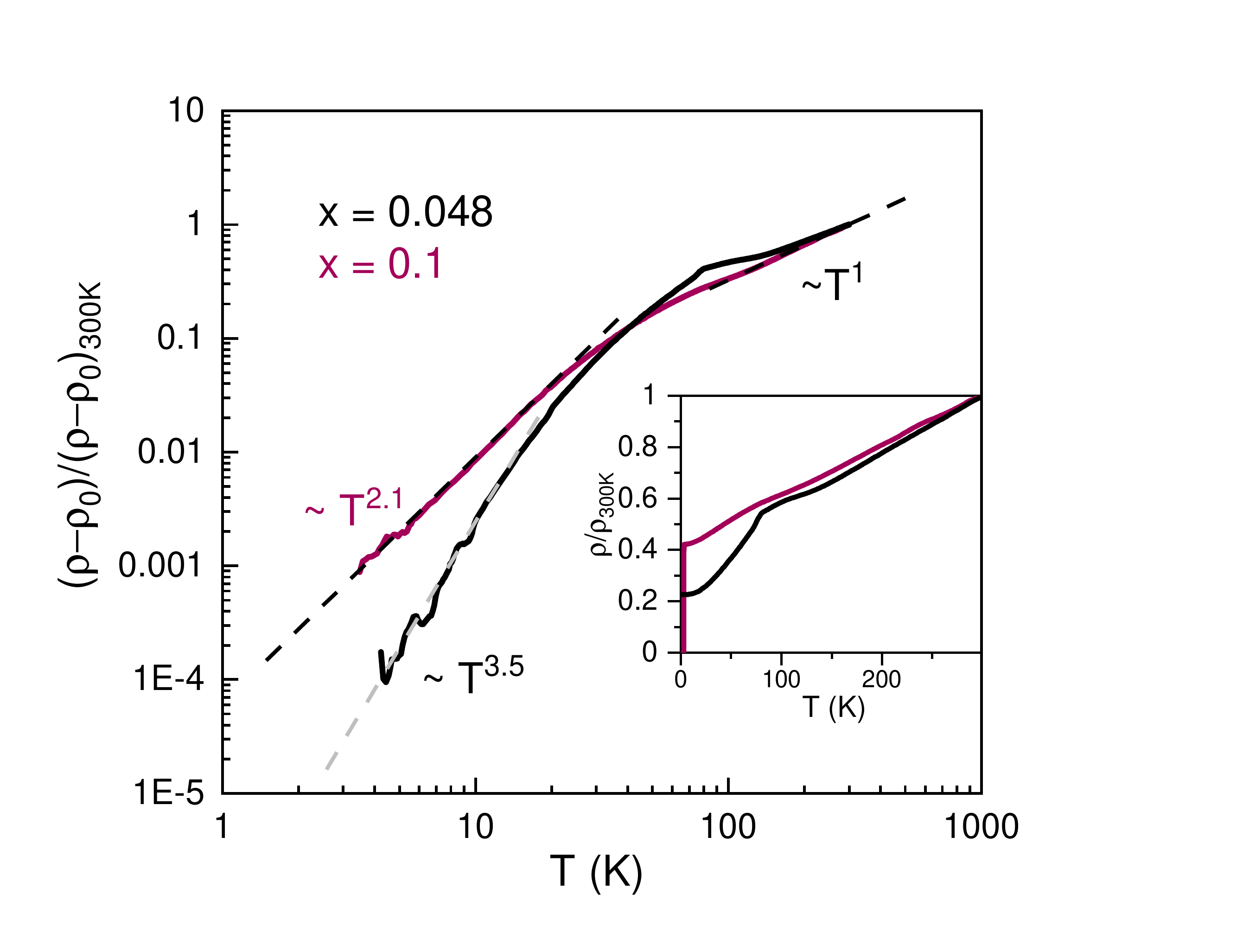}
\end{minipage}
\caption{Low-temperature power-law dependence of the resistivity for x = 0.048 and x = 0.1. The residual resistivity, $\rho_0$ has been subtracted and the data were then normalized at 300 K.  The inset displays  $\rho / \rho_\textrm{300K}$ versus T. }
\label{Fig9}
\end{figure}

Finally, we compare the resistivity of the high- and low-T$_c$ crystals in Fig. \ref{Fig9}. Here, in order to examine the low-T power law behavior, we plot the normalized $\rho - \rho_0$ versus temperature on a log-log plot.  Interestingly, the resistivity of the higher-T$_c$ crystal (x = 0.1) scales like T$^{\textrm{2.1}}$, suggestive of electron-electron scattering.  In contrast we find T$^{\textrm{3.5}}$ for the low-T$_c$ crystal (x = 0.048), which is closer to what is expected for electron-phonon scattering, i.e. T$^{\textrm{5}}$ .  At high temperature, both sets of data roughly approach a linear temperature dependence.

\section{\label{sec:level1}Discussion}

The present results provide compelling evidence that the phase diagram of BaNi$_2$(As,P)$_2$ is far more interesting and intriguing than suggested by the original dat of Kudo et al~\cite{KudoPRL2012}.  We interpret the four-fold symmetry breaking transition at T$_\textrm{orth}$ as the long-range-order of the I-CDW~\cite{MerzPRB2021}, and the triclinic phase as a kind of lock-in transition, i.e. C-CDW order.  The dominant structural change of the unit cell parameters accompanying these transitions is a large reduction of the c/a ratio in the tetragonal setting, which thus provides an effective tuning parameter in this system. Both the I-CDW and the C-CDW transitions are most likely driven by the same physical mechanism, since c/a evolves in a monotonic fashion upon crossing both transitions.

The I-CDW transition is overshadowed by a large high-temperature fluctuating region, as evidenced by our thermal-expansion data and resulting temperature dependence of the c/a ratio.  We note that the onset temperature of these fluctuations, T$_\textrm{fluc}$, is roughly coincident with the observation of the E$_{g,1}$ phonon broadening/splitting observed in Raman data ~\cite{Yao2022}. T$_\textrm{cross}$ on the other hand, roughly corresponds to the marked increase of the I-CDW peak intensity in x-ray diffraction \cite{MerzPRB2021,Yao2022}.  A weak signal of the I-CDW can in fact already be observed at room temperature \cite{Merzpriv}, indicating a very large fluctuation region. The large difference between the onset of the I-CDW signal at T$_\textrm{cross}$ and the actual phase transition at T$_\textrm{orth}$ again is also consistent with a highly fluctuating order. Near-edge x-ray absorption fine structure (NEXAFS) data ~\cite{MerzPRB2021} show that charge/orbital fluctuations exist far above the phase transition and that a temperature-induced charge transfer from d$_{xy,yz}$ to d$_{xy}$ orbitals occurs, which likely is related to the presently observed large changes in c/a. The peak of the B$_\textrm{1g}$ elastoresistivity signal occurs at T$_\textrm{orth}$, and the signal is mostly confined to the region between T$_\textrm{orth}$ and T$_\textrm{cross}$  ~\cite{EckergNP2020,Fracchet2022}.  These large fluctuations likely result from the low-dimensional nature of the CDW, which forms quasi-1D chains of staggered Ni atoms along one of the original $\left[ 100 \right]_{\textrm{}}$ tetragonal axes~\cite{KudoPRL2012,SefatPRB2009,LeePRL2019,MerzPRB2021}.  The here suggested low-dimensional physics is also consistent with the fact that there is very little entropy change associated with the long-range ordering transition at T$_\textrm{orth}$ (see Fig. 2b). In this sense, the actual symmetry breaking at T$_\textrm{orth}$ can be considered a minor secondary effect.

Since the I-CDW superlattice peaks persist to higher P-substitution levels into the higher-T$_c$ phase, the transition at T$_x$ might constitute a different kind of long-range order of the I-CDW than the orthorhombic structure.  More detailed structural investigations are needed to clarify this.  Although clearly visible in the thermal-expansion data, this transition appears quite subtle, since no clear signatures have been observed in heat capacity, resistivity, or Raman so far ~\cite{KudoPRL2012,Fracchet2022,Yao2022}. Elastoresistivity data exhibit only a broad maximum at 50 K ~\cite{Fracchet2022}, which roughly matches the transition at T$_x$, although the data do not show any significant hysteresis. 

Next we address the increase of the superconducting critical temperature with P-substitution. First, we point out that fully P-substituted BaNi$_2$As$_2$, i.e. BaNi$_2$P$_2$, is already superconducting with a T$_\textrm{c}$ of roughly 2.5 K - 3 K  ~\cite{JohrendtJSSC1997,MineSSC2008}, which raises the question of whether fluctuations of the triclinic phase are necessary for producing the higher-T$_\textrm{c}$ at x = 0.1, or whether it is more appropriate to think of a T$_\textrm{c}$ suppression due to the C-CDW transition and associated triclinic distortion.  On the other hand, similar to BaNi$_2$(As,P)$_2$, superconductivity in BaNi$_2$(P,Ge)$_2$  is strongest (i.e. highest T$_\textrm{c}$ and coupling strength $\Delta C_\textrm{p}/\gamma T_\textrm{c}$) near the onset of a structural distortion or charge-density wave  ~\cite{HiraiPRB2012}). In this system, a phonon softening was also observed at this transition.  An increase in electron-phonon coupling strength, $\lambda$, is usually reflected in an increase in $\gamma$ = $N_{bare}(1+\lambda)$, which is however not observed in either of these systems \cite{KudoPRL2012,HiraiPRB2012}. This apparent contradiction might be reconciled by the recent theoretical work using renormalized Migdal-Eliashberg theory \cite{NosarzewskiPRB2021}. The strong anharmonicity of the low-lying phonons, as evidenced by the presently observed extremely large Grüneisen parameters, as well as the extremely broad low-lying E$_{g,1}$ mode \cite{Yao2022}, could contribute to the stronger coupling \cite{SettyPRB2020,BaggioliPRL2019}.

\section{\label{sec:level1}Conclusions}

In conclusion, we have demonstrated that the phase diagram of BaNi$_2$(As,P)$_2$ is much richer than suggested by the original data of Kudo et al. ~\cite{KudoPRL2012}, hosting a variety of structural phase transitions related to charge/orbital ordering.   Significant precursors above the I-CDW and C-CDW transitions are seen in the anisotropic thermal expansion and c/a ratio of the crystal structure, pointing to low-dimensionality and large fluctations of the charge/orbital ordering.
The higher-T$_c$ material is characterized by an additional phase transition to an hitherto unknown structure and by an extremely anharmonic phonon mode, which is likely related to the very broad low-lying E$_{g,1}$ mode observed by Raman\cite{Yao2022}. Associated with this mode are unusually large and diverging Grüneisen parameters, pointing to a possible novel phonon quantum criticality.  Finally, it would be very interesting to perform similar thermal-expansion and shear-modulus measurements on the closely related (Ba,Sr)Ni$_2$As$_2$ system, in which the increase of T$_\textrm{c}$ has been attributed to electronic nematicity \cite{EckergNP2020}.

\begin{acknowledgments}

Work at KIT was partially funded by the Deutsche Forschungsgemeinschaft (DFG, German Research Foundation) - TRR 288-422213477 (projects A02 and B03). A.S. acknowledges funding from the European Union’s Horizon 2020 research and innovation program under the Marie Skłdowska-Curie grant agreement No. 847471 (QUSTEC). KW acknowledges funding from the Swiss National Science Foundation through the Postdoc mobility program. 
\end{acknowledgments}





\bibliographystyle{apsrev4-2}	
\bibliography{BaNi2AsP2CDW.bib}

\begin{thebibliography}{38}%
\makeatletter
\providecommand \@ifxundefined [1]{%
 \@ifx{#1\undefined}
}%
\providecommand \@ifnum [1]{%
 \ifnum #1\expandafter \@firstoftwo
 \else \expandafter \@secondoftwo
 \fi
}%
\providecommand \@ifx [1]{%
 \ifx #1\expandafter \@firstoftwo
 \else \expandafter \@secondoftwo
 \fi
}%
\providecommand \natexlab [1]{#1}%
\providecommand \enquote  [1]{``#1''}%
\providecommand \bibnamefont  [1]{#1}%
\providecommand \bibfnamefont [1]{#1}%
\providecommand \citenamefont [1]{#1}%
\providecommand \href@noop [0]{\@secondoftwo}%
\providecommand \href [0]{\begingroup \@sanitize@url \@href}%
\providecommand \@href[1]{\@@startlink{#1}\@@href}%
\providecommand \@@href[1]{\endgroup#1\@@endlink}%
\providecommand \@sanitize@url [0]{\catcode `\\12\catcode `\$12\catcode
  `\&12\catcode `\#12\catcode `\^12\catcode `\_12\catcode `\%12\relax}%
\providecommand \@@startlink[1]{}%
\providecommand \@@endlink[0]{}%
\providecommand \url  [0]{\begingroup\@sanitize@url \@url }%
\providecommand \@url [1]{\endgroup\@href {#1}{\urlprefix }}%
\providecommand \urlprefix  [0]{URL }%
\providecommand \Eprint [0]{\href }%
\providecommand \doibase [0]{https://doi.org/}%
\providecommand \selectlanguage [0]{\@gobble}%
\providecommand \bibinfo  [0]{\@secondoftwo}%
\providecommand \bibfield  [0]{\@secondoftwo}%
\providecommand \translation [1]{[#1]}%
\providecommand \BibitemOpen [0]{}%
\providecommand \bibitemStop [0]{}%
\providecommand \bibitemNoStop [0]{.\EOS\space}%
\providecommand \EOS [0]{\spacefactor3000\relax}%
\providecommand \BibitemShut  [1]{\csname bibitem#1\endcsname}%
\let\auto@bib@innerbib\@empty
\bibitem [{\citenamefont {Pfisterer}\ and\ \citenamefont
  {Nagorsen}(1980)}]{PfistererNagorsen1980}%
  \BibitemOpen
  \bibfield  {author} {\bibinfo {author} {\bibfnamefont {M.}~\bibnamefont
  {Pfisterer}}\ and\ \bibinfo {author} {\bibfnamefont {G.}~\bibnamefont
  {Nagorsen}},\ }\href {https://doi.org/doi:10.1515/znb-1980-0611} {\bibfield
  {journal} {\bibinfo  {journal} {Zeitschrift für Naturforschung B}\ }\textbf
  {\bibinfo {volume} {35}},\ \bibinfo {pages} {703} (\bibinfo {year}
  {1980})}\BibitemShut {NoStop}%
\bibitem [{\citenamefont {Ronning}\ \emph {et~al.}(2008)\citenamefont
  {Ronning}, \citenamefont {Kurita}, \citenamefont {Bauer}, \citenamefont
  {Scott}, \citenamefont {Park}, \citenamefont {Klimczuk}, \citenamefont
  {Movshovich},\ and\ \citenamefont {Thompson}}]{Ronning_2008}%
  \BibitemOpen
  \bibfield  {author} {\bibinfo {author} {\bibfnamefont {F.}~\bibnamefont
  {Ronning}}, \bibinfo {author} {\bibfnamefont {N.}~\bibnamefont {Kurita}},
  \bibinfo {author} {\bibfnamefont {E.~D.}\ \bibnamefont {Bauer}}, \bibinfo
  {author} {\bibfnamefont {B.~L.}\ \bibnamefont {Scott}}, \bibinfo {author}
  {\bibfnamefont {T.}~\bibnamefont {Park}}, \bibinfo {author} {\bibfnamefont
  {T.}~\bibnamefont {Klimczuk}}, \bibinfo {author} {\bibfnamefont
  {R.}~\bibnamefont {Movshovich}},\ and\ \bibinfo {author} {\bibfnamefont
  {J.~D.}\ \bibnamefont {Thompson}},\ }\href
  {https://doi.org/10.1088/0953-8984/20/34/342203} {\bibfield  {journal}
  {\bibinfo  {journal} {Journal of Physics: Condensed Matter}\ }\textbf
  {\bibinfo {volume} {20}},\ \bibinfo {pages} {342203} (\bibinfo {year}
  {2008})}\BibitemShut {NoStop}%
\bibitem [{\citenamefont {Sefat}\ \emph {et~al.}(2009)\citenamefont {Sefat},
  \citenamefont {McGuire}, \citenamefont {Jin}, \citenamefont {Sales},
  \citenamefont {Mandrus}, \citenamefont {Ronning}, \citenamefont {Bauer},\
  and\ \citenamefont {Mozharivskyj}}]{SefatPRB2009}%
  \BibitemOpen
  \bibfield  {author} {\bibinfo {author} {\bibfnamefont {A.~S.}\ \bibnamefont
  {Sefat}}, \bibinfo {author} {\bibfnamefont {M.~A.}\ \bibnamefont {McGuire}},
  \bibinfo {author} {\bibfnamefont {R.}~\bibnamefont {Jin}}, \bibinfo {author}
  {\bibfnamefont {B.~C.}\ \bibnamefont {Sales}}, \bibinfo {author}
  {\bibfnamefont {D.}~\bibnamefont {Mandrus}}, \bibinfo {author} {\bibfnamefont
  {F.}~\bibnamefont {Ronning}}, \bibinfo {author} {\bibfnamefont {E.~D.}\
  \bibnamefont {Bauer}},\ and\ \bibinfo {author} {\bibfnamefont
  {Y.}~\bibnamefont {Mozharivskyj}},\ }\href
  {https://doi.org/10.1103/PhysRevB.79.094508} {\bibfield  {journal} {\bibinfo
  {journal} {Phys. Rev. B}\ }\textbf {\bibinfo {volume} {79}},\ \bibinfo
  {pages} {094508} (\bibinfo {year} {2009})}\BibitemShut {NoStop}%
\bibitem [{\citenamefont {Kothapalli}\ \emph {et~al.}(2010)\citenamefont
  {Kothapalli}, \citenamefont {Ronning}, \citenamefont {Bauer}, \citenamefont
  {Schultz},\ and\ \citenamefont {Nakotte}}]{Kothapalli}%
  \BibitemOpen
  \bibfield  {author} {\bibinfo {author} {\bibfnamefont {K.}~\bibnamefont
  {Kothapalli}}, \bibinfo {author} {\bibfnamefont {F.}~\bibnamefont {Ronning}},
  \bibinfo {author} {\bibfnamefont {E.~D.}\ \bibnamefont {Bauer}}, \bibinfo
  {author} {\bibfnamefont {A.~J.}\ \bibnamefont {Schultz}},\ and\ \bibinfo
  {author} {\bibfnamefont {H.}~\bibnamefont {Nakotte}},\ }\href@noop {}
  {\bibfield  {journal} {\bibinfo  {journal} {Journal of Physics: Conference
  Series}\ }\textbf {\bibinfo {volume} {251}},\ \bibinfo {pages} {012010}
  (\bibinfo {year} {2010})}\BibitemShut {NoStop}%
\bibitem [{\citenamefont {Zhou}\ \emph {et~al.}(2011)\citenamefont {Zhou},
  \citenamefont {Xu}, \citenamefont {Zhang}, \citenamefont {Xu}, \citenamefont
  {He}, \citenamefont {Yang}, \citenamefont {Chen}, \citenamefont {Xie},
  \citenamefont {Cui}, \citenamefont {Arita}, \citenamefont {Shimada},
  \citenamefont {Namatame}, \citenamefont {Taniguchi}, \citenamefont {Dai},\
  and\ \citenamefont {Feng}}]{zhouPRB2011}%
  \BibitemOpen
  \bibfield  {author} {\bibinfo {author} {\bibfnamefont {B.}~\bibnamefont
  {Zhou}}, \bibinfo {author} {\bibfnamefont {M.}~\bibnamefont {Xu}}, \bibinfo
  {author} {\bibfnamefont {Y.}~\bibnamefont {Zhang}}, \bibinfo {author}
  {\bibfnamefont {G.}~\bibnamefont {Xu}}, \bibinfo {author} {\bibfnamefont
  {C.}~\bibnamefont {He}}, \bibinfo {author} {\bibfnamefont {L.~X.}\
  \bibnamefont {Yang}}, \bibinfo {author} {\bibfnamefont {F.}~\bibnamefont
  {Chen}}, \bibinfo {author} {\bibfnamefont {B.~P.}\ \bibnamefont {Xie}},
  \bibinfo {author} {\bibfnamefont {X.-Y.}\ \bibnamefont {Cui}}, \bibinfo
  {author} {\bibfnamefont {M.}~\bibnamefont {Arita}}, \bibinfo {author}
  {\bibfnamefont {K.}~\bibnamefont {Shimada}}, \bibinfo {author} {\bibfnamefont
  {H.}~\bibnamefont {Namatame}}, \bibinfo {author} {\bibfnamefont
  {M.}~\bibnamefont {Taniguchi}}, \bibinfo {author} {\bibfnamefont
  {X.}~\bibnamefont {Dai}},\ and\ \bibinfo {author} {\bibfnamefont {D.~L.}\
  \bibnamefont {Feng}},\ }\href {https://doi.org/10.1103/PhysRevB.83.035110}
  {\bibfield  {journal} {\bibinfo  {journal} {Phys. Rev. B}\ }\textbf {\bibinfo
  {volume} {83}},\ \bibinfo {pages} {035110} (\bibinfo {year}
  {2011})}\BibitemShut {NoStop}%
\bibitem [{\citenamefont {Eckberg}\ \emph {et~al.}(2018)\citenamefont
  {Eckberg}, \citenamefont {Wang}, \citenamefont {Hodovanets}, \citenamefont
  {Kim}, \citenamefont {Campbell}, \citenamefont {Zavalij}, \citenamefont
  {Piccoli},\ and\ \citenamefont {Paglione}}]{EckergPRB2018}%
  \BibitemOpen
  \bibfield  {author} {\bibinfo {author} {\bibfnamefont {C.}~\bibnamefont
  {Eckberg}}, \bibinfo {author} {\bibfnamefont {L.}~\bibnamefont {Wang}},
  \bibinfo {author} {\bibfnamefont {H.}~\bibnamefont {Hodovanets}}, \bibinfo
  {author} {\bibfnamefont {H.}~\bibnamefont {Kim}}, \bibinfo {author}
  {\bibfnamefont {D.~J.}\ \bibnamefont {Campbell}}, \bibinfo {author}
  {\bibfnamefont {P.}~\bibnamefont {Zavalij}}, \bibinfo {author} {\bibfnamefont
  {P.}~\bibnamefont {Piccoli}},\ and\ \bibinfo {author} {\bibfnamefont
  {J.}~\bibnamefont {Paglione}},\ }\href
  {https://doi.org/10.1103/PhysRevB.97.224505} {\bibfield  {journal} {\bibinfo
  {journal} {Phys. Rev. B}\ }\textbf {\bibinfo {volume} {97}},\ \bibinfo
  {pages} {224505} (\bibinfo {year} {2018})}\BibitemShut {NoStop}%
\bibitem [{\citenamefont {Lee}\ \emph {et~al.}(2019)\citenamefont {Lee},
  \citenamefont {Goh},\ and\ \citenamefont {Cho}}]{LeePRL2019}%
  \BibitemOpen
  \bibfield  {author} {\bibinfo {author} {\bibfnamefont {S.-H.}\ \bibnamefont
  {Lee}}, \bibinfo {author} {\bibfnamefont {J.~S.}\ \bibnamefont {Goh}},\ and\
  \bibinfo {author} {\bibfnamefont {D.}~\bibnamefont {Cho}},\ }\href
  {https://doi.org/10.1103/PhysRevLett.122.106404} {\bibfield  {journal}
  {\bibinfo  {journal} {Phys. Rev. Lett.}\ }\textbf {\bibinfo {volume} {122}},\
  \bibinfo {pages} {106404} (\bibinfo {year} {2019})}\BibitemShut {NoStop}%
\bibitem [{\citenamefont {Lee}\ \emph {et~al.}(2021)\citenamefont {Lee},
  \citenamefont {Collini}, \citenamefont {Sun}, \citenamefont {Mitrano},
  \citenamefont {Guo}, \citenamefont {Eckberg}, \citenamefont {Paglione},
  \citenamefont {Fradkin},\ and\ \citenamefont {Abbamonte}}]{LeePRL2021}%
  \BibitemOpen
  \bibfield  {author} {\bibinfo {author} {\bibfnamefont {S.}~\bibnamefont
  {Lee}}, \bibinfo {author} {\bibfnamefont {J.}~\bibnamefont {Collini}},
  \bibinfo {author} {\bibfnamefont {S.~X.-L.}\ \bibnamefont {Sun}}, \bibinfo
  {author} {\bibfnamefont {M.}~\bibnamefont {Mitrano}}, \bibinfo {author}
  {\bibfnamefont {X.}~\bibnamefont {Guo}}, \bibinfo {author} {\bibfnamefont
  {C.}~\bibnamefont {Eckberg}}, \bibinfo {author} {\bibfnamefont
  {J.}~\bibnamefont {Paglione}}, \bibinfo {author} {\bibfnamefont
  {E.}~\bibnamefont {Fradkin}},\ and\ \bibinfo {author} {\bibfnamefont
  {P.}~\bibnamefont {Abbamonte}},\ }\href
  {https://doi.org/10.1103/PhysRevLett.127.027602} {\bibfield  {journal}
  {\bibinfo  {journal} {Phys. Rev. Lett.}\ }\textbf {\bibinfo {volume} {127}},\
  \bibinfo {pages} {027602} (\bibinfo {year} {2021})}\BibitemShut {NoStop}%
\bibitem [{\citenamefont {Merz}\ \emph {et~al.}(2021)\citenamefont {Merz},
  \citenamefont {Wang}, \citenamefont {Wolf}, \citenamefont {Nagel},
  \citenamefont {Meingast},\ and\ \citenamefont {Schuppler}}]{MerzPRB2021}%
  \BibitemOpen
  \bibfield  {author} {\bibinfo {author} {\bibfnamefont {M.}~\bibnamefont
  {Merz}}, \bibinfo {author} {\bibfnamefont {L.}~\bibnamefont {Wang}}, \bibinfo
  {author} {\bibfnamefont {T.}~\bibnamefont {Wolf}}, \bibinfo {author}
  {\bibfnamefont {P.}~\bibnamefont {Nagel}}, \bibinfo {author} {\bibfnamefont
  {C.}~\bibnamefont {Meingast}},\ and\ \bibinfo {author} {\bibfnamefont
  {S.}~\bibnamefont {Schuppler}},\ }\bibfield  {journal} {\bibinfo  {journal}
  {Physical Review B}\ }\textbf {\bibinfo {volume} {104}},\ \href
  {https://doi.org/10.1103/physrevb.104.184509} {10.1103/physrevb.104.184509}
  (\bibinfo {year} {2021})\BibitemShut {NoStop}%
\bibitem [{\citenamefont {Kudo}\ \emph {et~al.}(2012)\citenamefont {Kudo},
  \citenamefont {Takasuga}, \citenamefont {Okamoto}, \citenamefont {Hiroi},\
  and\ \citenamefont {Nohara}}]{KudoPRL2012}%
  \BibitemOpen
  \bibfield  {author} {\bibinfo {author} {\bibfnamefont {K.}~\bibnamefont
  {Kudo}}, \bibinfo {author} {\bibfnamefont {M.}~\bibnamefont {Takasuga}},
  \bibinfo {author} {\bibfnamefont {Y.}~\bibnamefont {Okamoto}}, \bibinfo
  {author} {\bibfnamefont {Z.}~\bibnamefont {Hiroi}},\ and\ \bibinfo {author}
  {\bibfnamefont {M.}~\bibnamefont {Nohara}},\ }\href
  {https://doi.org/10.1103/PhysRevLett.109.097002} {\bibfield  {journal}
  {\bibinfo  {journal} {Phys. Rev. Lett.}\ }\textbf {\bibinfo {volume} {109}},\
  \bibinfo {pages} {097002} (\bibinfo {year} {2012})}\BibitemShut {NoStop}%
\bibitem [{\citenamefont {Eckberg}\ \emph {et~al.}(2020)\citenamefont
  {Eckberg}, \citenamefont {Campbell}, \citenamefont {Metz}, \citenamefont
  {Collini}, \citenamefont {Hodovanets}, \citenamefont {Drye}, \citenamefont
  {Zavalij}, \citenamefont {Christensen}, \citenamefont {Fernandes},
  \citenamefont {Lee}, \citenamefont {Abbamonte}, \citenamefont {Lynn},\ and\
  \citenamefont {Paglione}}]{EckergNP2020}%
  \BibitemOpen
  \bibfield  {author} {\bibinfo {author} {\bibfnamefont {C.}~\bibnamefont
  {Eckberg}}, \bibinfo {author} {\bibfnamefont {D.~J.}\ \bibnamefont
  {Campbell}}, \bibinfo {author} {\bibfnamefont {T.}~\bibnamefont {Metz}},
  \bibinfo {author} {\bibfnamefont {J.}~\bibnamefont {Collini}}, \bibinfo
  {author} {\bibfnamefont {H.}~\bibnamefont {Hodovanets}}, \bibinfo {author}
  {\bibfnamefont {T.}~\bibnamefont {Drye}}, \bibinfo {author} {\bibfnamefont
  {P.}~\bibnamefont {Zavalij}}, \bibinfo {author} {\bibfnamefont {M.~H.}\
  \bibnamefont {Christensen}}, \bibinfo {author} {\bibfnamefont {R.~M.}\
  \bibnamefont {Fernandes}}, \bibinfo {author} {\bibfnamefont {S.}~\bibnamefont
  {Lee}}, \bibinfo {author} {\bibfnamefont {P.}~\bibnamefont {Abbamonte}},
  \bibinfo {author} {\bibfnamefont {J.~W.}\ \bibnamefont {Lynn}},\ and\
  \bibinfo {author} {\bibfnamefont {J.}~\bibnamefont {Paglione}},\ }\href
  {https://doi.org/10.1038/s41567-019-0736-9} {\bibfield  {journal} {\bibinfo
  {journal} {Nature Physics}\ }\textbf {\bibinfo {volume} {16}},\ \bibinfo
  {pages} {346} (\bibinfo {year} {2020})}\BibitemShut {NoStop}%
\bibitem [{\citenamefont {Noda}\ \emph {et~al.}(2017)\citenamefont {Noda},
  \citenamefont {Kudo}, \citenamefont {Takasuga}, \citenamefont {Nohara},
  \citenamefont {Sugimoto}, \citenamefont {Ootsuki}, \citenamefont {Kobayashi},
  \citenamefont {Horiba}, \citenamefont {Ono}, \citenamefont {Kumigashira},
  \citenamefont {Fujimori}, \citenamefont {Saini},\ and\ \citenamefont
  {Mizokawa}}]{NodaJPSJ2017}%
  \BibitemOpen
  \bibfield  {author} {\bibinfo {author} {\bibfnamefont {T.}~\bibnamefont
  {Noda}}, \bibinfo {author} {\bibfnamefont {K.}~\bibnamefont {Kudo}}, \bibinfo
  {author} {\bibfnamefont {M.}~\bibnamefont {Takasuga}}, \bibinfo {author}
  {\bibfnamefont {M.}~\bibnamefont {Nohara}}, \bibinfo {author} {\bibfnamefont
  {T.}~\bibnamefont {Sugimoto}}, \bibinfo {author} {\bibfnamefont
  {D.}~\bibnamefont {Ootsuki}}, \bibinfo {author} {\bibfnamefont
  {M.}~\bibnamefont {Kobayashi}}, \bibinfo {author} {\bibfnamefont
  {K.}~\bibnamefont {Horiba}}, \bibinfo {author} {\bibfnamefont
  {K.}~\bibnamefont {Ono}}, \bibinfo {author} {\bibfnamefont {H.}~\bibnamefont
  {Kumigashira}}, \bibinfo {author} {\bibfnamefont {A.}~\bibnamefont
  {Fujimori}}, \bibinfo {author} {\bibfnamefont {N.~L.}\ \bibnamefont
  {Saini}},\ and\ \bibinfo {author} {\bibfnamefont {T.}~\bibnamefont
  {Mizokawa}},\ }\href {https://doi.org/10.7566/JPSJ.86.064708} {\bibfield
  {journal} {\bibinfo  {journal} {Journal of the Physical Society of Japan}\
  }\textbf {\bibinfo {volume} {86}},\ \bibinfo {pages} {064708} (\bibinfo
  {year} {2017})},\ \Eprint
  {https://arxiv.org/abs/https://doi.org/10.7566/JPSJ.86.064708}
  {https://doi.org/10.7566/JPSJ.86.064708} \BibitemShut {NoStop}%
\bibitem [{\citenamefont {Yamakawa}\ \emph {et~al.}(2013)\citenamefont
  {Yamakawa}, \citenamefont {Onari},\ and\ \citenamefont
  {Kontani}}]{YamakawaJPSJ2013}%
  \BibitemOpen
  \bibfield  {author} {\bibinfo {author} {\bibfnamefont {Y.}~\bibnamefont
  {Yamakawa}}, \bibinfo {author} {\bibfnamefont {S.}~\bibnamefont {Onari}},\
  and\ \bibinfo {author} {\bibfnamefont {H.}~\bibnamefont {Kontani}},\ }\href
  {https://doi.org/10.7566/JPSJ.82.094704} {\bibfield  {journal} {\bibinfo
  {journal} {Journal of the Physical Society of Japan}\ }\textbf {\bibinfo
  {volume} {82}},\ \bibinfo {pages} {094704} (\bibinfo {year} {2013})},\
  \Eprint {https://arxiv.org/abs/https://doi.org/10.7566/JPSJ.82.094704}
  {https://doi.org/10.7566/JPSJ.82.094704} \BibitemShut {NoStop}%
\bibitem [{\citenamefont {Subedi}\ and\ \citenamefont
  {Singh}(2008)}]{SubediPRB2008}%
  \BibitemOpen
  \bibfield  {author} {\bibinfo {author} {\bibfnamefont {A.}~\bibnamefont
  {Subedi}}\ and\ \bibinfo {author} {\bibfnamefont {D.~J.}\ \bibnamefont
  {Singh}},\ }\href {https://doi.org/10.1103/PhysRevB.78.132511} {\bibfield
  {journal} {\bibinfo  {journal} {Phys. Rev. B}\ }\textbf {\bibinfo {volume}
  {78}},\ \bibinfo {pages} {132511} (\bibinfo {year} {2008})}\BibitemShut
  {NoStop}%
\bibitem [{\citenamefont {B\"ohmer}\ \emph {et~al.}(2015)\citenamefont
  {B\"ohmer}, \citenamefont {Arai}, \citenamefont {Hardy}, \citenamefont
  {Hattori}, \citenamefont {Iye}, \citenamefont {Wolf}, \citenamefont
  {L\"ohneysen}, \citenamefont {Ishida},\ and\ \citenamefont
  {Meingast}}]{BohmerPRL2015}%
  \BibitemOpen
  \bibfield  {author} {\bibinfo {author} {\bibfnamefont {A.~E.}\ \bibnamefont
  {B\"ohmer}}, \bibinfo {author} {\bibfnamefont {T.}~\bibnamefont {Arai}},
  \bibinfo {author} {\bibfnamefont {F.}~\bibnamefont {Hardy}}, \bibinfo
  {author} {\bibfnamefont {T.}~\bibnamefont {Hattori}}, \bibinfo {author}
  {\bibfnamefont {T.}~\bibnamefont {Iye}}, \bibinfo {author} {\bibfnamefont
  {T.}~\bibnamefont {Wolf}}, \bibinfo {author} {\bibfnamefont {H.~v.}\
  \bibnamefont {L\"ohneysen}}, \bibinfo {author} {\bibfnamefont
  {K.}~\bibnamefont {Ishida}},\ and\ \bibinfo {author} {\bibfnamefont
  {C.}~\bibnamefont {Meingast}},\ }\href
  {https://doi.org/10.1103/PhysRevLett.114.027001} {\bibfield  {journal}
  {\bibinfo  {journal} {Phys. Rev. Lett.}\ }\textbf {\bibinfo {volume} {114}},\
  \bibinfo {pages} {027001} (\bibinfo {year} {2015})}\BibitemShut {NoStop}%
\bibitem [{\citenamefont {Yao}\ and\ \citenamefont {etc}()}]{Yao2022}%
  \BibitemOpen
  \bibfield  {author} {\bibinfo {author} {\bibfnamefont {Y.}~\bibnamefont
  {Yao}}\ and\ \bibinfo {author} {\bibnamefont {etc}},\ }\href@noop {}
  {\bibinfo  {journal} {to be published}\ }\BibitemShut {NoStop}%
\bibitem [{\citenamefont {Frachet}\ \emph {et~al.}(2022)\citenamefont
  {Frachet}, \citenamefont {Wiecki}, \citenamefont {Lacmann}, \citenamefont
  {Souliou}, \citenamefont {Willa}, \citenamefont {Meingast}, \citenamefont
  {Merz}, \citenamefont {Haghighirad}, \citenamefont {Tacon},\ and\
  \citenamefont {Böhmer}}]{Fracchet2022}%
  \BibitemOpen
\bibfield  {journal} {  }\bibfield  {author} {\bibinfo {author} {\bibfnamefont
  {M.}~\bibnamefont {Frachet}}, \bibinfo {author} {\bibfnamefont {P.~W.}\
  \bibnamefont {Wiecki}}, \bibinfo {author} {\bibfnamefont {T.}~\bibnamefont
  {Lacmann}}, \bibinfo {author} {\bibfnamefont {S.~M.}\ \bibnamefont
  {Souliou}}, \bibinfo {author} {\bibfnamefont {K.}~\bibnamefont {Willa}},
  \bibinfo {author} {\bibfnamefont {C.}~\bibnamefont {Meingast}}, \bibinfo
  {author} {\bibfnamefont {M.}~\bibnamefont {Merz}}, \bibinfo {author}
  {\bibfnamefont {A.~A.}\ \bibnamefont {Haghighirad}}, \bibinfo {author}
  {\bibfnamefont {M.~L.}\ \bibnamefont {Tacon}},\ and\ \bibinfo {author}
  {\bibfnamefont {A.~E.}\ \bibnamefont {Böhmer}},\ }\href
  {https://doi.org/10.48550/ARXIV.2207.02462} {\bibinfo {title}
  {Elastoresistivity in the incommensurate charge density wave phase of
  bani$_{\textrm{2} }$(as$_{\textrm{1-x}}$p$_{\textrm{x}}$)$_{\textrm{2}}$}}
  (\bibinfo {year} {2022})\BibitemShut {NoStop}%
\bibitem [{\citenamefont {Meingast}\ \emph {et~al.}(1990)\citenamefont
  {Meingast}, \citenamefont {Blank}, \citenamefont {B\"urkle}, \citenamefont
  {Obst}, \citenamefont {Wolf}, \citenamefont {W\"uhl}, \citenamefont
  {Selvamanickam},\ and\ \citenamefont {Salama}}]{ChristophPRB1990}%
  \BibitemOpen
  \bibfield  {author} {\bibinfo {author} {\bibfnamefont {C.}~\bibnamefont
  {Meingast}}, \bibinfo {author} {\bibfnamefont {B.}~\bibnamefont {Blank}},
  \bibinfo {author} {\bibfnamefont {H.}~\bibnamefont {B\"urkle}}, \bibinfo
  {author} {\bibfnamefont {B.}~\bibnamefont {Obst}}, \bibinfo {author}
  {\bibfnamefont {T.}~\bibnamefont {Wolf}}, \bibinfo {author} {\bibfnamefont
  {H.}~\bibnamefont {W\"uhl}}, \bibinfo {author} {\bibfnamefont
  {V.}~\bibnamefont {Selvamanickam}},\ and\ \bibinfo {author} {\bibfnamefont
  {K.}~\bibnamefont {Salama}},\ }\href
  {https://doi.org/10.1103/PhysRevB.41.11299} {\bibfield  {journal} {\bibinfo
  {journal} {Phys. Rev. B}\ }\textbf {\bibinfo {volume} {41}},\ \bibinfo
  {pages} {11299} (\bibinfo {year} {1990})}\BibitemShut {NoStop}%
\bibitem [{\citenamefont {B\"ohmer}\ \emph {et~al.}(2014)\citenamefont
  {B\"ohmer}, \citenamefont {Burger}, \citenamefont {Hardy}, \citenamefont
  {Wolf}, \citenamefont {Schweiss}, \citenamefont {Fromknecht}, \citenamefont
  {Reinecker}, \citenamefont {Schranz},\ and\ \citenamefont
  {Meingast}}]{BohmerPRL2014}%
  \BibitemOpen
  \bibfield  {author} {\bibinfo {author} {\bibfnamefont {A.~E.}\ \bibnamefont
  {B\"ohmer}}, \bibinfo {author} {\bibfnamefont {P.}~\bibnamefont {Burger}},
  \bibinfo {author} {\bibfnamefont {F.}~\bibnamefont {Hardy}}, \bibinfo
  {author} {\bibfnamefont {T.}~\bibnamefont {Wolf}}, \bibinfo {author}
  {\bibfnamefont {P.}~\bibnamefont {Schweiss}}, \bibinfo {author}
  {\bibfnamefont {R.}~\bibnamefont {Fromknecht}}, \bibinfo {author}
  {\bibfnamefont {M.}~\bibnamefont {Reinecker}}, \bibinfo {author}
  {\bibfnamefont {W.}~\bibnamefont {Schranz}},\ and\ \bibinfo {author}
  {\bibfnamefont {C.}~\bibnamefont {Meingast}},\ }\href
  {https://doi.org/10.1103/PhysRevLett.112.047001} {\bibfield  {journal}
  {\bibinfo  {journal} {Phys. Rev. Lett.}\ }\textbf {\bibinfo {volume} {112}},\
  \bibinfo {pages} {047001} (\bibinfo {year} {2014})}\BibitemShut {NoStop}%
\bibitem [{\citenamefont {Pokharel}\ \emph {et~al.}(2022)\citenamefont
  {Pokharel}, \citenamefont {Grigorev}, \citenamefont {Mejas}, \citenamefont
  {Dong}, \citenamefont {Haghighirad}, \citenamefont {Heid}, \citenamefont
  {Yao}, \citenamefont {Merz}, \citenamefont {Le~Tacon},\ and\ \citenamefont
  {Demsar}}]{Pokharel2022}%
  \BibitemOpen
  \bibfield  {author} {\bibinfo {author} {\bibfnamefont {A.~R.}\ \bibnamefont
  {Pokharel}}, \bibinfo {author} {\bibfnamefont {V.}~\bibnamefont {Grigorev}},
  \bibinfo {author} {\bibfnamefont {A.}~\bibnamefont {Mejas}}, \bibinfo
  {author} {\bibfnamefont {T.}~\bibnamefont {Dong}}, \bibinfo {author}
  {\bibfnamefont {A.~A.}\ \bibnamefont {Haghighirad}}, \bibinfo {author}
  {\bibfnamefont {R.}~\bibnamefont {Heid}}, \bibinfo {author} {\bibfnamefont
  {Y.}~\bibnamefont {Yao}}, \bibinfo {author} {\bibfnamefont {M.}~\bibnamefont
  {Merz}}, \bibinfo {author} {\bibfnamefont {M.}~\bibnamefont {Le~Tacon}},\
  and\ \bibinfo {author} {\bibfnamefont {J.}~\bibnamefont {Demsar}},\ }\href
  {https://doi.org/10.1038/s42005-022-00919-x} {\bibfield  {journal} {\bibinfo
  {journal} {Communications Physics}\ }\textbf {\bibinfo {volume} {5}},\
  \bibinfo {pages} {141} (\bibinfo {year} {2022})}\BibitemShut {NoStop}%
\bibitem [{\citenamefont {Meingast}\ \emph {et~al.}(2012)\citenamefont
  {Meingast}, \citenamefont {Hardy}, \citenamefont {Heid}, \citenamefont
  {Adelmann}, \citenamefont {B\"ohmer}, \citenamefont {Burger}, \citenamefont
  {Ernst}, \citenamefont {Fromknecht}, \citenamefont {Schweiss},\ and\
  \citenamefont {Wolf}}]{Meingast2012PRL}%
  \BibitemOpen
  \bibfield  {author} {\bibinfo {author} {\bibfnamefont {C.}~\bibnamefont
  {Meingast}}, \bibinfo {author} {\bibfnamefont {F.}~\bibnamefont {Hardy}},
  \bibinfo {author} {\bibfnamefont {R.}~\bibnamefont {Heid}}, \bibinfo {author}
  {\bibfnamefont {P.}~\bibnamefont {Adelmann}}, \bibinfo {author}
  {\bibfnamefont {A.}~\bibnamefont {B\"ohmer}}, \bibinfo {author}
  {\bibfnamefont {P.}~\bibnamefont {Burger}}, \bibinfo {author} {\bibfnamefont
  {D.}~\bibnamefont {Ernst}}, \bibinfo {author} {\bibfnamefont
  {R.}~\bibnamefont {Fromknecht}}, \bibinfo {author} {\bibfnamefont
  {P.}~\bibnamefont {Schweiss}},\ and\ \bibinfo {author} {\bibfnamefont
  {T.}~\bibnamefont {Wolf}},\ }\href
  {https://doi.org/10.1103/PhysRevLett.108.177004} {\bibfield  {journal}
  {\bibinfo  {journal} {Phys. Rev. Lett.}\ }\textbf {\bibinfo {volume} {108}},\
  \bibinfo {pages} {177004} (\bibinfo {year} {2012})}\BibitemShut {NoStop}%
\bibitem [{\citenamefont {Wang}\ \emph {et~al.}(2019)\citenamefont {Wang},
  \citenamefont {He}, \citenamefont {Scherer}, \citenamefont {Hardy},
  \citenamefont {Schweiss}, \citenamefont {Wolf}, \citenamefont {Merz},
  \citenamefont {Andersen},\ and\ \citenamefont {Meingast}}]{Wang2019JPSJ}%
  \BibitemOpen
  \bibfield  {author} {\bibinfo {author} {\bibfnamefont {L.}~\bibnamefont
  {Wang}}, \bibinfo {author} {\bibfnamefont {M.}~\bibnamefont {He}}, \bibinfo
  {author} {\bibfnamefont {D.~D.}\ \bibnamefont {Scherer}}, \bibinfo {author}
  {\bibfnamefont {F.}~\bibnamefont {Hardy}}, \bibinfo {author} {\bibfnamefont
  {P.}~\bibnamefont {Schweiss}}, \bibinfo {author} {\bibfnamefont
  {T.}~\bibnamefont {Wolf}}, \bibinfo {author} {\bibfnamefont {M.}~\bibnamefont
  {Merz}}, \bibinfo {author} {\bibfnamefont {B.~M.}\ \bibnamefont {Andersen}},\
  and\ \bibinfo {author} {\bibfnamefont {C.}~\bibnamefont {Meingast}},\ }\href
  {https://doi.org/10.7566/JPSJ.88.104710} {\bibfield  {journal} {\bibinfo
  {journal} {Journal of the Physical Society of Japan}\ }\textbf {\bibinfo
  {volume} {88}},\ \bibinfo {pages} {104710} (\bibinfo {year} {2019})},\
  \Eprint {https://arxiv.org/abs/https://doi.org/10.7566/JPSJ.88.104710}
  {https://doi.org/10.7566/JPSJ.88.104710} \BibitemShut {NoStop}%
\bibitem [{\citenamefont {Böhmer}\ \emph {et~al.}(2015)\citenamefont
  {Böhmer}, \citenamefont {Hardy}, \citenamefont {Wang}, \citenamefont {Wolf},
  \citenamefont {Schweiss},\ and\ \citenamefont {Meingast}}]{BohmerNC2015}%
  \BibitemOpen
  \bibfield  {author} {\bibinfo {author} {\bibfnamefont {A.~E.}\ \bibnamefont
  {Böhmer}}, \bibinfo {author} {\bibfnamefont {F.}~\bibnamefont {Hardy}},
  \bibinfo {author} {\bibfnamefont {L.}~\bibnamefont {Wang}}, \bibinfo {author}
  {\bibfnamefont {T.}~\bibnamefont {Wolf}}, \bibinfo {author} {\bibfnamefont
  {P.}~\bibnamefont {Schweiss}},\ and\ \bibinfo {author} {\bibfnamefont
  {C.}~\bibnamefont {Meingast}},\ }\href {https://doi.org/10.1038/ncomms8911}
  {\bibfield  {journal} {\bibinfo  {journal} {Nature Communications}\ }\textbf
  {\bibinfo {volume} {6}},\ \bibinfo {pages} {7911} (\bibinfo {year}
  {2015})}\BibitemShut {NoStop}%
\bibitem [{\citenamefont {Wang}\ \emph {et~al.}(2016)\citenamefont {Wang},
  \citenamefont {Hardy}, \citenamefont {B\"ohmer}, \citenamefont {Wolf},
  \citenamefont {Schweiss},\ and\ \citenamefont {Meingast}}]{LiranPRB2016}%
  \BibitemOpen
  \bibfield  {author} {\bibinfo {author} {\bibfnamefont {L.}~\bibnamefont
  {Wang}}, \bibinfo {author} {\bibfnamefont {F.}~\bibnamefont {Hardy}},
  \bibinfo {author} {\bibfnamefont {A.~E.}\ \bibnamefont {B\"ohmer}}, \bibinfo
  {author} {\bibfnamefont {T.}~\bibnamefont {Wolf}}, \bibinfo {author}
  {\bibfnamefont {P.}~\bibnamefont {Schweiss}},\ and\ \bibinfo {author}
  {\bibfnamefont {C.}~\bibnamefont {Meingast}},\ }\href
  {https://doi.org/10.1103/PhysRevB.93.014514} {\bibfield  {journal} {\bibinfo
  {journal} {Phys. Rev. B}\ }\textbf {\bibinfo {volume} {93}},\ \bibinfo
  {pages} {014514} (\bibinfo {year} {2016})}\BibitemShut {NoStop}%
\bibitem [{\citenamefont {Rotter}\ \emph {et~al.}(2008)\citenamefont {Rotter},
  \citenamefont {Tegel}, \citenamefont {Johrendt}, \citenamefont
  {Schellenberg}, \citenamefont {Hermes},\ and\ \citenamefont
  {P\"ottgen}}]{Rotter2008PRB}%
  \BibitemOpen
  \bibfield  {author} {\bibinfo {author} {\bibfnamefont {M.}~\bibnamefont
  {Rotter}}, \bibinfo {author} {\bibfnamefont {M.}~\bibnamefont {Tegel}},
  \bibinfo {author} {\bibfnamefont {D.}~\bibnamefont {Johrendt}}, \bibinfo
  {author} {\bibfnamefont {I.}~\bibnamefont {Schellenberg}}, \bibinfo {author}
  {\bibfnamefont {W.}~\bibnamefont {Hermes}},\ and\ \bibinfo {author}
  {\bibfnamefont {R.}~\bibnamefont {P\"ottgen}},\ }\href
  {https://doi.org/10.1103/PhysRevB.78.020503} {\bibfield  {journal} {\bibinfo
  {journal} {Phys. Rev. B}\ }\textbf {\bibinfo {volume} {78}},\ \bibinfo
  {pages} {020503} (\bibinfo {year} {2008})}\BibitemShut {NoStop}%
\bibitem [{\citenamefont {Chu}\ \emph {et~al.}(2009)\citenamefont {Chu},
  \citenamefont {Analytis}, \citenamefont {Kucharczyk},\ and\ \citenamefont
  {Fisher}}]{ChuPRB2009}%
  \BibitemOpen
  \bibfield  {author} {\bibinfo {author} {\bibfnamefont {J.-H.}\ \bibnamefont
  {Chu}}, \bibinfo {author} {\bibfnamefont {J.~G.}\ \bibnamefont {Analytis}},
  \bibinfo {author} {\bibfnamefont {C.}~\bibnamefont {Kucharczyk}},\ and\
  \bibinfo {author} {\bibfnamefont {I.~R.}\ \bibnamefont {Fisher}},\ }\href
  {https://doi.org/10.1103/PhysRevB.79.014506} {\bibfield  {journal} {\bibinfo
  {journal} {Phys. Rev. B}\ }\textbf {\bibinfo {volume} {79}},\ \bibinfo
  {pages} {014506} (\bibinfo {year} {2009})}\BibitemShut {NoStop}%
\bibitem [{\citenamefont {Jongh}\ and\ \citenamefont
  {A.R.Miedema}(2001)}]{DeJonghAdvainPhyP2001}%
  \BibitemOpen
  \bibfield  {author} {\bibinfo {author} {\bibfnamefont {L.}~\bibnamefont
  {Jongh}}\ and\ \bibinfo {author} {\bibnamefont {A.R.Miedema}},\ }\href
  {https://doi.org/10.1080/00018730110101412} {\bibfield  {journal} {\bibinfo
  {journal} {Advances in Physics}\ }\textbf {\bibinfo {volume} {50}},\ \bibinfo
  {pages} {947} (\bibinfo {year} {2001})},\ \Eprint
  {https://arxiv.org/abs/https://doi.org/10.1080/00018730110101412}
  {https://doi.org/10.1080/00018730110101412} \BibitemShut {NoStop}%
\bibitem [{\citenamefont {Garst}\ and\ \citenamefont
  {Rosch}(2005)}]{GarstPRB2005}%
  \BibitemOpen
  \bibfield  {author} {\bibinfo {author} {\bibfnamefont {M.}~\bibnamefont
  {Garst}}\ and\ \bibinfo {author} {\bibfnamefont {A.}~\bibnamefont {Rosch}},\
  }\href {https://doi.org/10.1103/PhysRevB.72.205129} {\bibfield  {journal}
  {\bibinfo  {journal} {Phys. Rev. B}\ }\textbf {\bibinfo {volume} {72}},\
  \bibinfo {pages} {205129} (\bibinfo {year} {2005})}\BibitemShut {NoStop}%
\bibitem [{\citenamefont {Sachdev}\ and\ \citenamefont
  {Keimer}(2011)}]{SachdevPhyToday2011}%
  \BibitemOpen
  \bibfield  {author} {\bibinfo {author} {\bibfnamefont {S.}~\bibnamefont
  {Sachdev}}\ and\ \bibinfo {author} {\bibfnamefont {B.}~\bibnamefont
  {Keimer}},\ }\href {https://doi.org/10.1063/1.3554314} {\bibfield  {journal}
  {\bibinfo  {journal} {Physics Today}\ }\textbf {\bibinfo {volume} {64}},\
  \bibinfo {pages} {29} (\bibinfo {year} {2011})},\ \Eprint
  {https://arxiv.org/abs/https://doi.org/10.1063/1.3554314}
  {https://doi.org/10.1063/1.3554314} \BibitemShut {NoStop}%
\bibitem [{\citenamefont {Wang}\ \emph {et~al.}(2018)\citenamefont {Wang},
  \citenamefont {He}, \citenamefont {Hardy}, \citenamefont {Adelmann},
  \citenamefont {Wolf}, \citenamefont {Merz}, \citenamefont {Schweiss},\ and\
  \citenamefont {Meingast}}]{LiranPRB2018}%
  \BibitemOpen
  \bibfield  {author} {\bibinfo {author} {\bibfnamefont {L.}~\bibnamefont
  {Wang}}, \bibinfo {author} {\bibfnamefont {M.}~\bibnamefont {He}}, \bibinfo
  {author} {\bibfnamefont {F.}~\bibnamefont {Hardy}}, \bibinfo {author}
  {\bibfnamefont {P.}~\bibnamefont {Adelmann}}, \bibinfo {author}
  {\bibfnamefont {T.}~\bibnamefont {Wolf}}, \bibinfo {author} {\bibfnamefont
  {M.}~\bibnamefont {Merz}}, \bibinfo {author} {\bibfnamefont {P.}~\bibnamefont
  {Schweiss}},\ and\ \bibinfo {author} {\bibfnamefont {C.}~\bibnamefont
  {Meingast}},\ }\href {https://doi.org/10.1103/PhysRevB.97.224518} {\bibfield
  {journal} {\bibinfo  {journal} {Phys. Rev. B}\ }\textbf {\bibinfo {volume}
  {97}},\ \bibinfo {pages} {224518} (\bibinfo {year} {2018})}\BibitemShut
  {NoStop}%
\bibitem [{\citenamefont {Guo}\ \emph {et~al.}(2022)\citenamefont {Guo},
  \citenamefont {Klemm}, \citenamefont {Oh}, \citenamefont {Xie}, \citenamefont
  {Lei}, \citenamefont {Gorovikov}, \citenamefont {Pedersen}, \citenamefont
  {Michiardi}, \citenamefont {Zhdanovich}, \citenamefont {Damascelli},
  \citenamefont {Denlinger}, \citenamefont {Hashimoto}, \citenamefont {Lu},
  \citenamefont {Mo}, \citenamefont {Moore}, \citenamefont {Birgeneau},
  \citenamefont {Singh}, \citenamefont {Dai},\ and\ \citenamefont
  {Yi}}]{Guo2022}%
  \BibitemOpen
  \bibfield  {author} {\bibinfo {author} {\bibfnamefont {Y.}~\bibnamefont
  {Guo}}, \bibinfo {author} {\bibfnamefont {M.}~\bibnamefont {Klemm}}, \bibinfo
  {author} {\bibfnamefont {J.~S.}\ \bibnamefont {Oh}}, \bibinfo {author}
  {\bibfnamefont {Y.}~\bibnamefont {Xie}}, \bibinfo {author} {\bibfnamefont
  {B.-H.}\ \bibnamefont {Lei}}, \bibinfo {author} {\bibfnamefont
  {S.}~\bibnamefont {Gorovikov}}, \bibinfo {author} {\bibfnamefont
  {T.}~\bibnamefont {Pedersen}}, \bibinfo {author} {\bibfnamefont
  {M.}~\bibnamefont {Michiardi}}, \bibinfo {author} {\bibfnamefont
  {S.}~\bibnamefont {Zhdanovich}}, \bibinfo {author} {\bibfnamefont
  {A.}~\bibnamefont {Damascelli}}, \bibinfo {author} {\bibfnamefont
  {J.}~\bibnamefont {Denlinger}}, \bibinfo {author} {\bibfnamefont
  {M.}~\bibnamefont {Hashimoto}}, \bibinfo {author} {\bibfnamefont
  {D.}~\bibnamefont {Lu}}, \bibinfo {author} {\bibfnamefont {S.-K.}\
  \bibnamefont {Mo}}, \bibinfo {author} {\bibfnamefont {R.~G.}\ \bibnamefont
  {Moore}}, \bibinfo {author} {\bibfnamefont {R.~J.}\ \bibnamefont
  {Birgeneau}}, \bibinfo {author} {\bibfnamefont {D.~J.}\ \bibnamefont
  {Singh}}, \bibinfo {author} {\bibfnamefont {P.}~\bibnamefont {Dai}},\ and\
  \bibinfo {author} {\bibfnamefont {M.}~\bibnamefont {Yi}},\ }\href
  {https://doi.org/10.48550/ARXIV.2205.14339} {\bibinfo {title} {Spectral
  evidence for unidirectional charge density wave in detwinned bani$_2$as$_2$}}
  (\bibinfo {year} {2022})\BibitemShut {NoStop}%
\bibitem [{\citenamefont {Merz}()}]{Merzpriv}%
  \BibitemOpen
  \bibfield  {author} {\bibinfo {author} {\bibfnamefont {M.}~\bibnamefont
  {Merz}},\ }\href@noop {} {\bibinfo  {journal} {private communication}\
  }\BibitemShut {NoStop}%
\bibitem [{\citenamefont {Johrendt}\ \emph {et~al.}(1997)\citenamefont
  {Johrendt}, \citenamefont {Felser}, \citenamefont {Jepsen}, \citenamefont
  {Andersen}, \citenamefont {Mewis},\ and\ \citenamefont
  {Rouxel}}]{JohrendtJSSC1997}%
  \BibitemOpen
\bibfield  {journal} {  }\bibfield  {author} {\bibinfo {author} {\bibfnamefont
  {D.}~\bibnamefont {Johrendt}}, \bibinfo {author} {\bibfnamefont
  {C.}~\bibnamefont {Felser}}, \bibinfo {author} {\bibfnamefont
  {O.}~\bibnamefont {Jepsen}}, \bibinfo {author} {\bibfnamefont {O.~K.}\
  \bibnamefont {Andersen}}, \bibinfo {author} {\bibfnamefont {A.}~\bibnamefont
  {Mewis}},\ and\ \bibinfo {author} {\bibfnamefont {J.}~\bibnamefont
  {Rouxel}},\ }\href {https://doi.org/https://doi.org/10.1006/jssc.1997.7300}
  {\bibfield  {journal} {\bibinfo  {journal} {Journal of Solid State
  Chemistry}\ }\textbf {\bibinfo {volume} {130}},\ \bibinfo {pages} {254}
  (\bibinfo {year} {1997})}\BibitemShut {NoStop}%
\bibitem [{\citenamefont {Mine}\ \emph {et~al.}(2008)\citenamefont {Mine},
  \citenamefont {Yanagi}, \citenamefont {Kamiya}, \citenamefont {Kamihara},
  \citenamefont {Hirano},\ and\ \citenamefont {Hosono}}]{MineSSC2008}%
  \BibitemOpen
  \bibfield  {author} {\bibinfo {author} {\bibfnamefont {T.}~\bibnamefont
  {Mine}}, \bibinfo {author} {\bibfnamefont {H.}~\bibnamefont {Yanagi}},
  \bibinfo {author} {\bibfnamefont {T.}~\bibnamefont {Kamiya}}, \bibinfo
  {author} {\bibfnamefont {Y.}~\bibnamefont {Kamihara}}, \bibinfo {author}
  {\bibfnamefont {M.}~\bibnamefont {Hirano}},\ and\ \bibinfo {author}
  {\bibfnamefont {H.}~\bibnamefont {Hosono}},\ }\href
  {https://doi.org/https://doi.org/10.1016/j.ssc.2008.05.010} {\bibfield
  {journal} {\bibinfo  {journal} {Solid State Communications}\ }\textbf
  {\bibinfo {volume} {147}},\ \bibinfo {pages} {111} (\bibinfo {year}
  {2008})}\BibitemShut {NoStop}%
\bibitem [{\citenamefont {Hirai}\ \emph {et~al.}(2012)\citenamefont {Hirai},
  \citenamefont {von Rohr},\ and\ \citenamefont {Cava}}]{HiraiPRB2012}%
  \BibitemOpen
  \bibfield  {author} {\bibinfo {author} {\bibfnamefont {D.}~\bibnamefont
  {Hirai}}, \bibinfo {author} {\bibfnamefont {F.}~\bibnamefont {von Rohr}},\
  and\ \bibinfo {author} {\bibfnamefont {R.~J.}\ \bibnamefont {Cava}},\ }\href
  {https://doi.org/10.1103/PhysRevB.86.100505} {\bibfield  {journal} {\bibinfo
  {journal} {Phys. Rev. B}\ }\textbf {\bibinfo {volume} {86}},\ \bibinfo
  {pages} {100505} (\bibinfo {year} {2012})}\BibitemShut {NoStop}%
\bibitem [{\citenamefont {Nosarzewski}\ \emph {et~al.}(2021)\citenamefont
  {Nosarzewski}, \citenamefont {Sch\"uler},\ and\ \citenamefont
  {Devereaux}}]{NosarzewskiPRB2021}%
  \BibitemOpen
  \bibfield  {author} {\bibinfo {author} {\bibfnamefont {B.}~\bibnamefont
  {Nosarzewski}}, \bibinfo {author} {\bibfnamefont {M.}~\bibnamefont
  {Sch\"uler}},\ and\ \bibinfo {author} {\bibfnamefont {T.~P.}\ \bibnamefont
  {Devereaux}},\ }\href {https://doi.org/10.1103/PhysRevB.103.024520}
  {\bibfield  {journal} {\bibinfo  {journal} {Phys. Rev. B}\ }\textbf {\bibinfo
  {volume} {103}},\ \bibinfo {pages} {024520} (\bibinfo {year}
  {2021})}\BibitemShut {NoStop}%
\bibitem [{\citenamefont {Setty}\ \emph {et~al.}(2020)\citenamefont {Setty},
  \citenamefont {Baggioli},\ and\ \citenamefont {Zaccone}}]{SettyPRB2020}%
  \BibitemOpen
  \bibfield  {author} {\bibinfo {author} {\bibfnamefont {C.}~\bibnamefont
  {Setty}}, \bibinfo {author} {\bibfnamefont {M.}~\bibnamefont {Baggioli}},\
  and\ \bibinfo {author} {\bibfnamefont {A.}~\bibnamefont {Zaccone}},\ }\href
  {https://doi.org/10.1103/PhysRevB.102.174506} {\bibfield  {journal} {\bibinfo
   {journal} {Phys. Rev. B}\ }\textbf {\bibinfo {volume} {102}},\ \bibinfo
  {pages} {174506} (\bibinfo {year} {2020})}\BibitemShut {NoStop}%
\bibitem [{\citenamefont {Baggioli}\ and\ \citenamefont
  {Zaccone}(2019)}]{BaggioliPRL2019}%
  \BibitemOpen
  \bibfield  {author} {\bibinfo {author} {\bibfnamefont {M.}~\bibnamefont
  {Baggioli}}\ and\ \bibinfo {author} {\bibfnamefont {A.}~\bibnamefont
  {Zaccone}},\ }\href {https://doi.org/10.1103/PhysRevLett.122.145501}
  {\bibfield  {journal} {\bibinfo  {journal} {Phys. Rev. Lett.}\ }\textbf
  {\bibinfo {volume} {122}},\ \bibinfo {pages} {145501} (\bibinfo {year}
  {2019})}\BibitemShut {NoStop}%
\end{thebibliography}%

\end{document}